\documentclass[twocolumn,useAMS]{mn2e}
\usepackage{graphics,times}
\input{epsf}
\def\bvarphi{\mbox{\boldmath $\varphi$}}
\def\bphi{\mbox{\boldmath $\phi$}}
\newcommand{\beq}{\begin{equation}}
\newcommand{\eeq}{\end{equation}}
\def\la{\hbox{\raise.35ex\rlap{$<$}\lower.6ex\hbox{$\sim$}\ }}
\def\ga{\hbox{\raise.35ex\rlap{$>$}\lower.6ex\hbox{$\sim$}\ }}

\def\beq{\begin{equation}}
\def\eeq{\end{equation}}
\def\beqa{\begin{eqnarray}}
\def\eeqa{\end{eqnarray}}

\def\order#1{{\cal O}\left({#1}\right)}
\newcommand{\sfrac}[2]{\small \mbox{$\frac{#1}{#2}$}}
\title[On the stratorotational instability]
{On the stratorotational instability in the quasi-hydrostatic semi-geostrophic limit
\thanks{Research supported by
    the Israel Science Foundation, the Helen and Robert Asher
    Fund and the Technion Fund for the Promotion of Research}}
\author[O.M. Umurhan ]{O.M. Umurhan$^{1,2}$\thanks{Email: mumurhan@physics.technion.ac.il},
\\
$^{1}$Department of Physics, Technion-Israel Institute of
Technology, 32000 Haifa, Israel\\
$^{2}$Department of Astronomy, City College of San Francisco, San Francisco, CA 94112, USA}

\date{Accepted ------------. Received ------------------}

\pagerange{\pageref{firstpage}--\pageref{lastpage}} \pubyear{2005}

\begin{document}

\label{firstpage}

\maketitle

\begin{abstract}
The linear normal-mode stratorotational instability (SRI) is analytically reexamined
in the inviscid limit where
the length scales of horizontal disturbances
are large compared their vertical and radial counterparts.  Boundary
conditions different than channel walls are also considered.
This quasi-hydrostatic, semi-geostrophic (QHSG) approximation
allows one to examine the effect of a vertically varying Brunt-Vaisaila frequency, $N^2$.
It is found that the normal-mode instability persists when $N^2$ increases quadratically with
respect to the disc vertical coordinate.  However we also find that the SRI seems
to exist in this inviscid QHSG extreme only for channel wall conditions: when one
or both of the reflecting walls are removed there is no instability in the asymptotic
limit explored here.  It is also found that only exponential-type SRI modes (as defined by
Dubrulle et al. 2005) exist under these conditions.
These equations also admit non-normal mode behaviour.
Fixed Lagrangian pressure conditions on both radial boundaries predicts
there to be no normal mode behaviour in the QHSG limit.
The mathematical relationship between the results obtained here and that of the
classic Eady (1949) problem for baroclinic instability is drawn.
We conjecture as to the mathematical/physical
nature of the SRI.\par
The general linear problem, analyzed without approximation in the context
of the Boussinesq equations, admits a potential vorticity-like quantity that is
advectively conserved by the shear.  Its existence means that a continuous
spectrum \emph{is a generic feature of this system}.  It also implies that in places where the
Brunt-Vaisaila frequency becomes dominant the linearized flow may two-dimensionalize
by advectively conserving its vertical vorticity.

\end{abstract}

\label{Introduction}
The question whether or not
hydrodynamic activity can emerge in protoplanetary
discs is experiencing a Renaissance.
Given the absence of an inflection point in the basic Keplerian
flow of discs, it was natural for investigations, beginning in
the early 90's, to consider other physical effects as a possible
source of supercritical linear instabilities.  The MRI (Balbus, 2003)
an instability (non-conservative) involving the joint
interplay of rotation and magnetic effects,
has proven itself to be a viable linear mechanism
which could lead to globally sustained activity in discs. \par
However,
fresh analysis of simplified models of protoplanetary discs, like the
shearing sheet approximation
(Goldreich \& Lynden-Bell, 1965)
utilized in many hydrodynamic and magneto-hydrodynamic
investigations of circumstellar Keplerian discs,
have shown that a number of alternative
routes to long-term activity, both linear and nonlinear, can occur
for purely hydrodynamic disturbances.  Bracco et al (1999) demonstrated
that anticyclonic vortices can live for long periods of
time before ultimately decaying away
in viscous two-dimensional, global disc models
of incompressible flow
executing Keplerian motion.  This work has since
led to many other investigations
which consider an array of dynamical processes, sometimes even
linearly stable,
that could operate
in a purely hydrodynamic manner and these include (but not limited
to): large amplitude defects of vorticity in the flow (Li et al. 2000)
leading to inflection point "secondary" - instabilities,
steady forcing of 2D linearized global viscous disc disturbances
which have shown to lead to strong transiently growing
structures and patterns (Iounnou \& Kakouris, 2002).  Transient growth in 2D disturbances,
an effect primarily driven by the Orr-mechanism of vortex-tilting (Orr 1907)
and which is fundamentally a non-normal mode effect
(Schmid \& Henningson, 2001, Chagelishvilli et al. 2003, Tevzadze et al., 2003,
Yecko 2004) can lead to sustained nonlinear activity in 2D,
including robust and coherent anticyclones (Umurhan \& Regev 2004).\par
Recently, Barranco and Marcus (2005), considering an analogous set of equations
describing the shearing-sheet limit of a circumstellar disc,
showed that nonlinear disturbances in 3D, where the disc vertical direction
and effect of gravity
are included, lead to long-lived vortex structures with vorticity
pointing parallel to the direction of gravity: something akin
to steady anticyclones persisting away from the disc midplane.  This is a
striking feature because it implies that in these regimes the flow
behaves nearly two-dimensionally.  This goes against
implications of the shearing sheet simulations of Balbus, Hawley \& Stone (1996)
and  Hawley, Balbus \& Winters (1999) which have suggested that there is
slim possibility of sustained dynamics in purely hydrodynamic systems
subject to strong rotation and shear.
\par
The effect of buoyancy on disturbances has only recently been
reconsidered.
According to the Solberg-Hoiland criterion it seemed
that a sufficient condition for a naturally occurring non-magnetic
instability in
a disc is for there
 to be an adverse entropy gradient in the direction
normal to the disc plane.
\footnote{The acoustic instability of Papaloizou \& Pringle
(1984,1985) is not considered "natural" in this sense because it relies on
the presence of artificial boundary conditions.  See Section 5 for further discussion
on this matter.}
However
the work of Yavneh et al. (2002) has recently turned this wisdom on its head.
They demonstrated
that a linear stability analysis of the classic Taylor Couette problem
(that is, flow between two concentric rotating cylinders) in which
the fluid is stratified (in the direction parallel to the cylinder axes)
leads to normal-mode instability even if the stratification
is stable to buoyant oscillations.
\par
Dubrulle et al., 2005 (D05)
showed that 3D linearized normal-mode disturbances
of stratified local sections of a protoplanetary disc can manifest
this instability.  The effect, termed the {\em Stratorotational
Instability} (SRI), has been proposed as a mechanism,
free of non-conservative effects like magnetic field interactions,
which could actively operate in the disc, possibly lead to
turbulence (D05), and has been suggested to be
relevant in connection to the steady vortices observed in the simulations
of Barranco \& Marcus (2005)
(Shalybkov \& Rudiger, 2005).
\par
The SRI effect, as revealed by Yavneh et al (2000), Shalybkov \& Rudiger (2005)
and in D05, was considered for flows in which the radial boundaries of the
system are rigid: in other words, under boundary conditions in which there is
no normal flow across the bounding channel walls.\footnote{ Note, however,
that
D05 also consider stress-free and periodic boundaries in the radial directions.}
In order to affect a tractable analysis
the investigations thus far have
assumed that the disc vertical component of gravity, $g {\bf {\hat z}}$,
and the vertical gradient of the basic state temperature, $\partial_z T_b$
(or, equivalently, the basic state entropy gradient, $\partial_z S_b$)
are {\em constant} with respect to the disc vertical coordinate.
Although this may be a good approximation for atmospheric flows like
the Earth's  (Pedlosky, 1987),  this is not an accurate
representation of conditions near the midplane of a circumstellar disc.
\par
Furthermore,
the question as to what are reasonable radial boundary conditions
for shearing-sheet sections of Keplerian discs remains very open.
Shearing-sheet sections of Keplerian discs, centered about
the disc midplane, have both a $g$ and a $\partial_z T_b$ ($\partial_z S_b$),
whose product is proportional to the square of the
{\em Brunt-Vaisaila frequency}, $N^2$,
which varies quadratically with respect to the height from the midplane, $z$.
It is natural then to ask (i)
whether or not the SRI is a dynamical effect driven primarily by the reflecting
property of the system's radial boundaries and, (ii) whether or not it persists
under conditions in which the Brunt-Vaisaila frequency
varies with respect to the vertical coordinate - something one would expect
in a protoplanetary disc.
\par
A purpose of this study is to revisit, primarily via asymptotic means,
the properties of the SRI instability as discussed by D05.
The main motivation of this inquiry is to see whether or not
the no-flow radial boundary conditions employed by their study
and/or whether or not the assumption of the constancy of $N^2$
significantly alters the SRI effect uncovered by previous investigations.
\par
We would like to make it clear here that the inclusion of stratification
in quantities like the vertical component of gravity and the temperature/entropy
gradients mathematically results in a normal mode problem
who is primarily non-separable in the radial and vertical coordinates.
It is for this reason a general analysis of this physical scenario is quite
difficult for both analytical and numerical reasons.
\par
We approach these questions by reconsidering the SRI within
the context of two model equations describing a local
section of a circumstellar disc.  We will experiment with the
effects of differing boundary conditions.
\par
In Section 1, we introduce the first of the equations, the Large-Shearing Box (LSB),
which are the basic equations appropriate to shearing-sheet sections
of accretion discs
centered about their midplanes.
These are then used to motivate a second, simpler, model equation
set: the incompressible Boussinesq equations (BE)
in inviscid three dimensional rotating plane Couette flow (or {\em rpCf}, see Yecko, 2004).
This simpler set is mathematically equivalent to the set
studied, in the inviscid limit, in D05.
Boundary conditions are considered which share a common property
in the BE,  namely those that
cause the total disturbance energy,
called $E$, to change solely due to
the energy exchanged between disturbances and the shear through the
Reynolds stress term and not due to work done on the system from
outside.  This is achieved by enforcing periodicity in the azimuthal
direction: either
periodicity in the disc vertical direction
(when constant $N^2$ is assumed)
or zero normal (vertical) velocity fluctuations at
some fiducial vertical disc boundary (when $N^2$ varies with disc height).
There a number of radial boundary conditions which achieves these objective
and we investigate these:  (a)
no-normal flow at the radial boundaries, sometimes referred to in this
manuscript as "channel-wall conditions", (b) no Lagrangian
variation of the pressure on the moving radial boundaries, (c) or some
mixed combination of these two.
\par
In Section 2 normal mode solutions of the BE are considered.
First it is shown that there exists an advectively conserved quantity of the flow.
The conservation
of this quantity in certain limits implies that the {\em linearized} flow is nearly two-dimensional -
in that the vertical vorticity is the dominant quantity that is preserved
by the shear advected flow.
We then proceed towards a normal mode analysis
by initially
assuming the constancy of the Brunt-Vaisaila frequency, $N^2$.  We asymptotically analyze disturbances
whose radial ($x$) and vertical length scales are
dwarfed by comparison to the azimuthal scales ($y$) - a circumstance which
is refereed to as the {\em quasi-hydrostatic semi-geostrophic
approximation} (QHSG).
This limit predisposes
the resulting equations into admitting
simple analytical solutions.  This asymptotic limit shares many of
the qualities of the quasi-WKB analysis considered in D05.
Implementing the variety of boundary conditions enumerated above
shows that, in this limit, only the no normal (radial) flow boundary conditions admit unstable normal modes.
We also find that there is always a continuous spectrum in this problem, irrespective
of the radial boundary conditions,
and we briefly discuss some of its features.
The section is rounded out
by relaxing the constancy of $N^2$ and considering the
case where it varies quadratically with respect to the disc height in
this same QHSG limit.  There appears to be a potential vorticity-like
quantity which is advectively conserved in the linear limit.  Moreover,
it turns out that
it is possible to construct separable
solutions (in $x$ and $z$) for the normal mode problem.
The main conclusion from this is that
the normal mode stability behaviour is
{\em unaffected} by the stratification of $N^2$.
Unlike in D05, where the SRI appears for both
 "exponential"-modes and "oscillatory"-modes, the results
 of this section show that the SRI only occurs for "exponential"-modes.
\par
In Section 3 the same QHSG approximation is applied directly
to the LSB.  These yield, similarly, a conserved potential vorticity
like quantity but which now includes the effects of weak compressibility
and a finite soundspeed.\footnote{Note that incompressibility has the mathematical
effect of an infinite soundspeed.}
The equations for the normal-modes are also separable
here and it is found, again, that the stability behaviour appears
to be unaffected by the inclusion of stratification and weak
compressibility effects.  This has been analyzed only for the case
of constant soundspeed (i.e. constant disc background temperature profile).
In Section 4 we discuss these results
and conjecture as to the disappearance of the SRI in this inviscid QHSG limit.
\par
The main results of this paper can be summarized here: (a) the SRI is also present
in the LSB model equations, (b)
we find that in the inviscid-QHSG limit
only the doubly reflecting boundary conditions, i.e. no-normal flow at the
two radial boundaries, seem to admit unstable normal modes,
(c) the inclusion of stratification realistic for a disc
(again, in the QHSG
approximation)
does not alter the SRI as obtained in previous investigations,
and (d) there is an advectively conserved quantity in the BE
which has the character of a potential vorticity; its existence implies both
that this type of flow always has a continuous spectrum
associated with it and that there exists certain conditions,
including large Brunt-Vaisaila frequencies,
in which the underlying flow behaves two dimensionally by
advectively conserving the vertical component of its vorticity.
\par
Note that the terms {\em channel flow} and {\em no normal flow} are
used interchangeably in this manuscript.

\section{Equations and Boundary Conditions}
\subsection{The Large Shearing Box and Boussinesq Equations}
There are a few formal asymptotic derivations for the
set of equations appropriate to the dynamics of a localized section
of a rotationally supported disc (e.g. Goldreich \& Lynden-Bell,1965).  For
this discussion we shall begin with the Large-Shearing Box (LSB)
equations as they appear in Umurhan \& Regev (2004),
\beqa
& & (\partial_t - q\Omega_0x \partial_y)\rho + \nabla\cdot (\rho_b+\rho)\bf u'
= 0, \label{lsb_continuity}\\
& & (\partial_t - q\Omega_0x \partial_y) u' + {\bf u'}
\cdot\nabla u' -2\Omega_0 v' = -\frac{\partial_x p}{\rho_b + \rho}
\label{lsb_full_radial}\\
%- \frac{\rho'g(z)}{\rho_b + \rho'}
& & (\partial_t - q\Omega_0x \partial_y) v' + {\bf u'}
\cdot\nabla v' + (2-q)\Omega_0u' = -\frac{\partial_y p}{\rho_b + \rho},
\label{lsb_azimuthal}\\
& & (\partial_t - q\Omega_0x \partial_y) w' + {\bf u'}
\cdot\nabla w'  = -\frac{\partial_z p}{\rho_b + \rho}
- \frac{\rho g(z)}{\rho_b + \rho},
\label{lsb_vertical}\\
%& & (\partial_t - q\Omega_0x \partial_y)p + {\bf u'}\cdot
%\nabla(p_b + p) + \gamma(p_b + p)\nabla \cdot {\bf u'} = 0.
& &(\partial_t - q\Omega_0x \partial_y)\Sigma
+ {\bf u'}\cdot \nabla \Sigma = 0
\label{lsb_entropy}
\eeqa
in which the total entropy is defined by
\[
\Sigma \equiv C_V \ln\frac{p_b + p}{\left(\rho_b + \rho\right)^\gamma}
\]
where $C_V$ is the specific heat at constant volume and where $\gamma$ is the usual
ratio of the specific heats at constant pressure and volumes.  \footnote{
In Umurhan \& Regev (2004), the entropy (heat) equation was written explicitly in terms
of the pressure and densities.  There is, then, no substantive difference between
these two expressions of the same conservation law}.
These equations represent the dynamics taking place in a "large-shearing
box" section near
the midplane of a Keplerian disc rotating with radius $R_0$ about the central star with the
local rotation vector,
$\Omega(R_0) {{\bf \hat z}}$.

The above equations have already been non-dimensionalized.  Time
is scaled by the local rotation time of the box.  All lengths have
been scaled according to a length scale $L$ which is
comparable to the disc thickness.  Pressures are
scaled according to the local scale of the soundspeed, which is in turn
based on some fiducial characteristic temperature scale.  For
further details  see Umurhan \& Regev (2004).  The above equations, in
which the vertical component of gravity is constant, are identical
to the equations considered by Tevzadze et al. (2003).

In the language of this paper,
$x$ corresponds to the shearwise (radial) coordinate which is
a small section located around the disc radius $R_0$, the azimuthal coordinate $y$ is streamwise
and $z$ is the vertical coordinate corresponding to the normal direction
of the original disc midplane.
The velocity disturbances in the radial, azimuthal and vertical directions
expressed by the variables ${\bf u'} = \{u',v',w'\}$.  These velocities represent
deviations over the steady Keplerian flow.

$\Omega_0$, sometimes also referred to as the  Coriolis parameter,
{\em is $1$ in these nondimensionalized units}, meaning to say because time has
been scaled according to the dimensional value of the rotation rate at $R_0$, i.e.
$\Omega(R_0)$, the local Coriolis parameter formally is equal to one.
We retain this symbol in
order to flag the Coriolis effects in this calculation.
The local shear gradient is defined to be
\beq
q \equiv -\left[\frac{R}{\Omega}\left(\frac{\partial \Omega}{\partial R}\right)\right]_{R_0},
\eeq
in which $\Omega(R)$ is the full disc rotation rate.
For Keplerian discs the value of $q$ is $3/2$.
The local Keplerian flow is represented here by
a linear shear in the azimuthal direction, i.e. $-q\Omega_0 x
{\bf \hat y}$.\par
The derivation of the LSB equations from the full-scale disc equations
begins from the observation that the disc is in fact rotationally supported
to leading order.
This means that in steady state the disc radial force balance is,
to leading order, between rotation and the radial component of gravity
as emanating from the central compact object.  The radial steady
state pressure gradients provide corrections which are on the order
of $\epsilon^2 = H^2/R_0^2$, where $H$ ia the characteristic height of the
disk.  Cold discs are those in which $\epsilon$ is formally assumed to be a small
quantity.
Given the scaling and expansion steps leading to the LSB this
rotational support translates to saying that a radial buoyancy
term in a rotationally supported disc is asymptotically smaller
than the Coriolis effect by $\order{\epsilon^2}$.  It is for this reason that there
is no buoyancy term in (\ref{lsb_full_radial}).
However, this rotational balance
implies that the {\em vertical structure} of the disc is primarily characterized
by the usual hydrostatic balance since,
by the symmetries inherent to the geometry of the problem, there
is no rotational support in that direction.  It is for this reason there is an
associated buoyancy term in (\ref{lsb_vertical}).
It should be noted that in the context of the LSB, the steady state
pressure and density functions are only functions of the $z$ coordinate (see below).
The vertically oriented gravitational field emanating from
the primary compact object is linearly proportional to the distance from the midplane
in the LSB system (again see next paragraph).
This is a consequence of the asymptotic expansions leading to the LSB
in which the central object's gravitational potential is expanded in
a Taylor series about the disc's symmetry plane (i.e. $z=0$).  Again, for
details of this procedure see Umurhan \& Regev (2004).
\par

The steady state density and
pressure functions are given by $p_b,\rho_b$.  These relate to each
other according to the aforementioned local hydrostatic equilibrium relationship,
\[
\partial_z p_b = -\rho_b g(z),
\]
with $g(z) = \Omega_0^2 z$.  The detailed solution for the steady states are then
determined once something has been said relating the steady quantities.
For simple theoretical investigations
this comes in the form of a barotropic equation of state, namely, the
statement that $p_b = p_b(\rho_b)$.  Note that in this paper, especially
in Section 4, we explicitly
assume that these steady quantities {\em depend only on $z$ and not on $x$}.
The dynamic pressures and densities,
$p$ and $\rho$, represent deviations about their corresponding steady
state quantities.
\par We want to gain
some insight as to the effects that gravity and gradients of state
quantities have on the local dynamics. We take an incremental
approach towards this goal by considering a more simplified
version of these equations.  In more concrete terms, the linear
theory (\ref{lsb_continuity}-\ref{lsb_entropy}) will show the
presence of three "types" of temporal modes, (i) a pair of
acoustic modes, (ii) a pair of inertial-gravity modes and (iii) an
entropy mode (Tevzadze et al., 2003). Although the acoustic modes
are interesting, we chose to consider the dynamics of a system in
which the acoustic modes are effectively filtered out.  \par To do
this we invoke the Boussinesq approximation which, in this sense,
we replace the continuity equation (\ref{lsb_continuity}) with the
statement of incompressibility and we replace the entropy
conservation equation (\ref{lsb_entropy}) with an evolution
equation for the temperature fluctuation, $\theta$.  All density
fluctuations are set to zero accept the one associated with the
buoyancy term in (\ref{lsb_continuity}) in which it is related to
the temperature fluctuation
 via
\[
\rho' = - \alpha_p \theta; \qquad \alpha_p \equiv
-\left({\partial \rho \over \partial T}\right)_p.
\]
In other words, $\alpha_p$ is the coefficient of thermal expansion at
constant pressure.  This is the typical formulation
of the Boussinesq approximation (Spiegel and Veronis, 1960).
We furthermore posit
that in this limit the basic state density profile is a constant
and in order to distinguish this from a spatially varying density profile
we denote the former with $\bar\rho_b$.  The resulting
model set of equations are similar to those assumed
in the studies by  Yavneh et al. (2001),
Dubrulle, et al. (2005) (these being viscous studies)
and Rudiger et al. (2005) (a cylindrical Taylor-Couette analysis).
We have then,
\beqa
& & \nabla\cdot \bf u' = 0, \label{gssb_incompressible}\\
& & (\partial_t - q\Omega_0x \partial_y) u' + {\bf u'}
\cdot\nabla u' -2\Omega_0 v' = -\frac{\partial_x p}{\bar\rho_b},
\label{gssb_radial}\\
%- \frac{\rho'g(z)}{\rho_b + \rho'}
& & (\partial_t - q\Omega_0x \partial_y) v' + {\bf u'}
\cdot\nabla v' + (2-q)\Omega_0u' = -\frac{\partial_y p}{\bar\rho_b},
\label{gssb_azimuthal}\\
& & (\partial_t - q\Omega_0x \partial_y) w' + {\bf u'}
\cdot\nabla w'  = -\frac{\partial_z p}{\bar\rho_b }
+ \frac{\theta g(z)\alpha_p}{\bar\rho_b},
\label{gssb_vertical}\\
& & (\partial_t - q\Omega_0x \partial_y)\theta' +
{\bf u'}\cdot \nabla\theta +
w \partial_z T_b = 0.
\label{gssb_theta}
\eeqa
These are otherwise known as the Boussinesq equations (BE) in plane-Couette shear.
The term $T_b$ is the way in which the basic state temperature profile
varies with height in this model formulation of the disc system.
Such a term would vary according to $\partial_z T_b = \bar T_{zz} z$, in
other words, the gradient of the basic state temperature has
a linear dependence with respect to the disc height in which
$\bar T_{zz}$ is the parameter that controls the slope of this variation.
For situations in which  $\bar T_{zz}$ is negative, the
atmosphere can be thought of as being classically
buoyantly unstable
which could lead to Rayleigh-Benard
convection (see for instance, Cabot, 1996).
\par
The BE equations here are mathematically
equivalent to the inviscid limit of the equations studied in D05.
The only difference here is in interpretation.  Whereas we follow a temperature
perturbation, $\theta$, and steady temperature profile, $T_b$, they follow an entropy
perturbation, denoted by $h$, and steady entropy distribution $H$.
The two disturbance quantities are related to each other via
\[
h = -\frac{\gamma \alpha_p C_{_V}}{\bar \rho_b} \theta.
\]
(Also see Appendix \ref{lsb_qhsg}).  $\gamma$ is the ratio of specific heats
and $C_{_V}$ is the specific heat at constant volume.
%Finally,
%aside from the equation for the temperature evolution,
%the set of equations (\ref{gssb_incompressible}-\ref{gssb_theta})
%are very similar to the equations for the "small-shearing box" as
%both derived and explored to some extent in Umurhan \& Regev (2004).
%We have added in, by hand, the effect of thermal effects in
%the form of Newton's Law of Cooling (Spiegel, 1957).
%$\Lambda$, which governs the cooling rate, is treated here as
%a constant and will be a tunable parameter. \par
The equations (\ref{gssb_incompressible}-\ref{gssb_theta})
are the basis of the discussion in Section 3. The
steady configuration of these equations which will be perturbed is
\beq
u'=v'=w'=\theta = 0, \qquad p = \bar p = {\rm constant}.
\label{steady_state}
\eeq

The set (\ref{lsb_continuity}-\ref{lsb_entropy}),
including their corresponding analogous boundary conditions will be considered
in Section 4.

\subsection{Boundary conditions}
There is no
obvious choice of boundary conditions for this reduced set of
inviscid flow equations.
We consider all disturbances to be periodic in the $y$ direction.  Because
these are equations meant to model what happens near the midplane of
a circumstellar disc, we can consider periodic conditions in the
vertical direction only under special circumstances (i.e.
the constant Brunt-Vaisaila frequency approximation of the BE
in Section 3, see below).
In a more general sense we will distinguish, instead,
between either varicose or sinuous modes.  By
{\em varicose modes} we mean to say disturbances which have even symmetry
with respect to the $z=0$ plane in all disturbances except the
vertical velocities, which have odd symmetry with respect
to the $z=0$ plane.  {\em Sinuous modes} have the reverse symmetry
of the varicose modes.  In situations in which a boundary condition needs
to be specified on vertical boundaries of the atmosphere, we assume
that there is no normal flow.
\par
The more troublesome of the boundary conditions
has to do with what to say about the flow variables in the radial direction.
This is because an injudicious choice of a boundary condition
might incite disturbances of the fluid into instability by drawing
energy across the boundaries.  It is our interest here to minimize
this potential as much as possible.
Of the myriad of possible choices, we see the following three sets of
boundary conditions as ones that achieve this objective
(and motivated further in Appendix \ref{integral_section}):
(a) that the flow be
confined between channel walls lying at $x = 0,1$, which means
in practice that the radial velocities are set to zero there, i.e., {\em
no normal-flow} conditions;
(b)
the flow has zero
{\em Lagrangian pressure} fluctuations
(defined below)
on both of the moving radial boundaries and;
(c) a mixture of these two conditions, for example,
by requiring there to be
no normal-flow on the inner boundary while
there is no Lagrangian pressure fluctuation at the outer boundary.
\par
The vanishing of the Lagrangian pressure fluctuation
on the undulating {\em radial} bounding surface ${\bf S_r}$ in linear theory translates to
requiring
\[
p' + \xi_x \partial_x \bar p = 0,
\]
where $p'$ is the pressure fluctuation about the steady state.
The position of any particular
radial surface, initially at rest at coordinate $x$, is denoted by
$\xi_x(y,z)$, and
evolves according to its Lagrangian equation of motion (Drazin \& Reid, 1984)
\beq
u'(x,y,z,t) =
\frac{d\xi_x}{dt} = (\partial_t - q\Omega_0 x \partial_y)\xi_x
+ v'\partial_y \xi_x + w\partial_z \xi_x.
\eeq
Because the steady state pressure configurations of both the BE ($\bar p$) and the LSB
($p_b$) are constant with respect to $x$,
the condition simplifies to requiring
\beq
p' = 0,
\eeq
at $x=0,1$.
%Although we do not do so here, one may also consider the vansishing of the
%radial derivative of the radial velocity at the domain walls, i.e. ,
%require that $\partial_x u' = 0$ and $x=0,1$.  The reason we do not consider
%this here is that it means, in general, that there may be an influx of energy
%across the domain walls according to the surface work term in
%(\ref{E_eval}).

\section{Linear Dynamics of the Boussinesq Equations}
In this inviscid limit there emerges a natural timescale defined
as the {\em Brunt-Vaisaila frequency}, $N$.  This time scale is
defined through the product of the vertical temperature gradient
and vertical gravity via \beq N^2 \equiv g \sfrac{1}{\bar
\rho_b}\alpha_p \partial_z T_b \label{def_BV_freq} \eeq which is,
in general, a function of the vertical coordinate $z$. Throughout
this study $N$ is taken to be real (buoyantly stable).
Linearization of (\ref{gssb_incompressible}-\ref{gssb_theta})
reduces to, \beqa \left(\partial_t -  q \Omega_0 x
\partial_y\right) u
 - 2\Omega_0 v &=& -\sfrac{1}{\bar \rho_b}
 \partial_x P,  \label{Nsc_u_lin}
\\
\left(\partial_t -  q \Omega_0 x \partial_y\right) v +
\Omega_0(2-q) u &=& -\sfrac{1}{\bar \rho_b}\partial_y P, \label{Nsc_v_lin}
\\
\left(\partial_t -  q \Omega_0 x \partial_y\right) w &=&
-\sfrac{1}{\bar \rho_b}\partial_z P + \Theta,
\label{Nsc_w_lin} \\
\left(\partial_t -  q \Omega_0 x \partial_y\right) \Theta
&=&  - N^2 w
, \label{Nsc_theta_lin} \\
\partial_x  u + \partial_y v + \partial_z w &=& 0,
\label{Nsc_incompressibility}
\eeqa
where the temperature variable has been slightly redefined as
$ \Theta \equiv g \alpha_p \theta/\bar \rho_b$.
$\bar\rho_b$  is set to $1$ from here on out.  It is now to be understood that unprimed
velocity expressions (i.e. $u,v,w$) represent linearized disturbances.
\subsection{A conserved quantity for linearized flow}
There exists a conserved quantity in these equations.  Operating
on (\ref{Nsc_u_lin}) by $\partial_y$ followed
by operating
(\ref{Nsc_v_lin}) by $\partial_x$ and subtracting the result reveals
\beq
\left(\partial_t -  q \Omega_0 x \partial_y\right)
\left(
\partial_x v - \partial_y u
\right) = \Omega_0(2-q) \partial_z w,
\eeq
where the incompressibility condition has been used.
The term on the LHS of this expression is the {\em
vertical vorticity}, i.e. $\zeta \equiv \partial_x v - \partial_y u$.
With a similar tack one can multiply (\ref{Nsc_theta_lin})
by $\Omega_0(2-q)/N^2$ and then operate on the result with
$\partial_z$ to get
\beq
\left(\partial_t -  q \Omega_0 x \partial_y\right)
\left(
\frac{\partial}{\partial z}\frac{\Omega_0(2-q)}{N^2}\Theta
\right) = -\Omega_0(2-q) \partial_z w.
\eeq
Adding the results together yields a general conserved quantity
of linearized flow of this type:
\beq
\left(\partial_t -  q \Omega_0 x \partial_x\right)\Xi
= 0,
\label{conserved_Xi}
\eeq
where
\beq \Xi =
\zeta +
\frac{\partial}{\partial z}\frac{\Omega_0(2-q)}{N^2}\Theta.
\label{def_Xi}
\eeq
The quantity $\Xi$ can be thought of as a generalized potential
vorticity for this type of flow whose analogous quantity is
discussed in Tevzadze et al. (2003).
The conservation of $\Xi$ immediately
implies that there always exists a continuous spectrum (see Schmid \&
Henningson, 2001, Tevzadze et al., 2003) for
this type of physical system.
Note also that
the system of linearized equations (i.e. \ref{Nsc_u_lin}-\ref{Nsc_incompressibility})
is third order in time.  One may suppose that there are
three independent normal modes for any given set of quantum numbers of the system
(see below), however,
given that there exists a conserved quantity, together with its
associated continuous spectrum, it means that there are at most only two normal modes
for any given quantum number set.\par
Inspection of $\Xi$ shows that disturbances behave in a quasi two-dimensional fashion
in some limits.  One of these is when $q=2$, that is at the critical Rayleigh
condition (Drazin \& Reid, 1984): it follows that
the vertical vorticity is conserved by the flow.  The second of these is to
notice that if the temperature fluctuation remains an order one quantity as
$N^2$ gets large then the flow again exhibits quasi two-dimensionality with
the vertical vorticity being conserved.  We reflect upon the consequences of this
conserved quantity some more in the Discussion.

\subsection{Constant $N^2$}\label{BE_equations_constantN2}
\par
Since part of the purpose of this work is to further develop some amount of intuition
about the dynamics of such disc environments
primarily through analytical means, it will be more tractable
for us to first treat the vertical gravity component $g(z)$ to be
\beq
g(z) = g_0{\rm sgn}(z).
\label{g_appx}
\eeq
The non-dimensional constant $g_0$ is technically arbitrary.
In a  similar vein we approximate the steady state temperature gradient by
saying
\beq
\partial_z T_b = \bar T_z {\rm sgn}(z),
\label{Tb_appx}
\eeq
in which $\bar T_z$ is another non-dimensional parameter.  The consequence of
this is that  $N^2$ is a constant for $z\neq 0$, and is zero at $z=0$.
At this stage, these assumptions are qualitatively no different than
what has be done in Yavneh et al. (2001), D05 and Shalybkov \& Rudiger (2005),
although we take a more realistic interpretation of a constant $N^2$ (and see
below).
%\par
%\subsection{Linearization and asymptotics}
From here on out we set $\bar \rho_b = 1$. Additionally, we
restrict analysis of the dynamics to $z>0$ and keeping in mind
that modes are considered to have either sinuous or varicose
spatial character in the vertical.
\par
We write general solutions into the form
\beq
\left(
\begin{array}{c}
u \\
v \\
P
\end{array}
\right)
=
\left(
\begin{array}{c}
u_{_{\alpha\beta}} \\
v_{_{\alpha\beta}} \\
P_{_{\alpha\beta}}
\end{array}
\right)
\left\{
\begin{array}{c}
\cos \beta z \\
\sin \beta z
\end{array}
\right\}
e^{i\omega t + i\alpha y}  + c.c.,
\label{solform_I} \eeq
while for the other variables
\beq
\left(
\begin{array}{c}
w \\
\Theta
\end{array}
\right)
=
\left(
\begin{array}{c}
w_{_{\alpha\beta}} \\
\Theta_{_{\alpha\beta}}
\end{array}
\right)
\left\{
\begin{array}{c}
-\sin \beta z \\
\cos \beta z
\end{array}
\right\}
e^{i\omega t + i\alpha y}  + c.c.,
\label{solform_II} \eeq
The terms above in the curly brackets represent varicose
disturbances while the terms below are the sinuous disturbances.
In this sense the "quantum numbers" of the system are given
by $\alpha$, $\beta$ (varicose or sinuous) and a radial overtone
number (if there are more than one) subject to solution of the normal mode boundary value problem
below.\footnote{The use of {\em quantum numbers} should be considered
only in terms of conventional nomenclature.  There is no real quantization in the horizontal
and vertical directions per se since we allow these quantities to take on
any value from the continuum of real numbers}.
Note that because this is a single Fourier expansion
we restrict our considerations to
$0<\alpha<\infty$ together with
$0 <  \beta < \infty$.
Insertion of (\ref{solform_I}-\ref{solform_II})
into the governing linear equations gives,
\beqa
i\left(\omega -  q \Omega_0 x \alpha\right)
u_{_{\alpha \beta}} - 2\Omega_0 v_{_{\alpha \beta}} &=&
-\partial_x P_{_{\alpha \beta}},  \label{nsc_u_lin}
\\
i\left(\omega - q \Omega_0 x \alpha\right)
v_{_{\alpha \beta}} + \Omega_0(2-q) u_{_{\alpha \beta}} &=&
-i\alpha P_{_{\alpha \beta}}, \label{nsc_v_lin}
\\
i\left(\omega - q \Omega_0 x \alpha\right) w_{_{\alpha \beta}}
&=&  -\beta  P_{_{\alpha \beta}}
- \Theta_{_{\alpha \beta}},
\label{nsc_w_lin} \\
i\left(\omega - q \Omega_0 x \alpha\right) \Theta_{_{\alpha \beta}}
&=&  - N^2 w_{_{\alpha \beta}}
, \label{nsc_theta_lin} \\
\partial_x  u_{_{\alpha \beta}} +i \alpha
v_{_{\alpha \beta}} - \beta  w_{_{\alpha \beta}} &=& 0,
\label{nsc_incompressibility}
\eeqa
\par

It turns
out that it is much more tractable to consider the linearized
normal-mode behavior in terms
of equations describing
the pressure fluctuation and radial velocities.  This
is entirely analogous to what was done in D05 and the following equations
should be compared to the ones quoted in D05 as Equations (21-22).
\beqa
\left(\frac{\beta^2\sigma^2}{N^2 - \sigma^2 }
- \alpha^2\right)
i P_{_{\alpha\beta}}
&=& -\sigma\partial_x  u_{_{\alpha\beta}} + \Omega_0(2-q)\alpha
  u_{_{\alpha\beta}} \nonumber \\
(\sigma^2 - \omega_\epsilon^2)  u_{_{\alpha\beta}} &=&
(\sigma\partial_x   iP_{_{\alpha\beta}} + 2\Omega_0\alpha   iP_{_{\alpha\beta}}),
\label{normal_mode_grpCf}
\eeqa
where, for the sake of
compact notation we use the
expression $\sigma \equiv \omega - q\Omega_0 \alpha x$.
The epicyclic frequency $\omega_\epsilon^2$ is
equivalent to the expression
$ 2(2-q)\Omega_0^2$.
\par
We are mainly interested in analytically expressible solutions to the
above set of equations.  To achieve this in an asymptotically
rigorous manner the following scalings seem natural:
when the horizontal wavenumber is
small, it follows that the frequency
scales similarly.
Using $\epsilon$ to measure this
smallness it follows,
\[
\alpha = \epsilon \alpha_1, \qquad \omega = \epsilon\omega_1 + \cdots.
\]
To lowest order it also follows that $\sigma = \epsilon\sigma_1 + \cdots$.
The pressure and velocities are consequently expanded by
\beqa
  P_{_{\alpha\beta}} &=& P_0 + \epsilon^2 P_2 + \cdots \nonumber \\
  u_{_{\alpha\beta}} &=& \epsilon u_1 + \epsilon^3 u_3 + \cdots \nonumber
\eeqa
Implementing these expansions into the governing equations
(\ref{normal_mode_grpCf}) yields at lowest order in $\epsilon$ a single equation
for the pressure perturbation,
\beq
(\omega_1 - q\Omega_0 \alpha_1 x)
\left(\partial_x^2  - F_e^2\beta^2\right) P_0
= 0.
\label{asymptotic_P0_eqn}
\eeq
%The equation for the pressure perturbation, in the asymptotic limit, simplifies to
%\beq
%(\omega_1 - q\Omega_0 \alpha_1 x)
%\left(\partial_x^2  - F_e^2\beta^2\right) P_0
%= 0.
%\label{asymptotic_P0_eqn_nocooling}
%\eeq
Normal-mode type
solutions to (\ref{asymptotic_P0_eqn}) are,
\beq
P_0 = A\cosh k_{_F}x + B\sinh k_{_F} x,
\eeq
where the {\em Froude-wavenumber}, $k_{_F}$,  is defined as
$k_{_F}^2 \equiv \omega_e^2 \beta^2/N^2
= \beta^2 F_e^2$. The {\em epicyclic-Froude number} is denoted
by $F_e$.
This mathematical structure of
(\ref{asymptotic_P0_eqn}) is identical to the operator
describing the evolution of plane-Couette disturbances in a channel
(e.g. Case, 1960).
\par
We compare the analytical solutions generated
here with numerical solutions generated for the boundary
value problem defined by the un-approximated full linearized equation
set (\ref{normal_mode_grpCf}).   A second order correct
(in the $x$ direction derivatives)
Newton-Raphson-Kantorovitch (NRK) relaxation scheme on a grid of
approximately 1000 to 2000 points is used for the verification.
Relative convergence was
checked by doubling the size of the domain.
Eigenvalues are determined
with errors that were no more than $\order{5\times 10^{-6}}$.  As
such the
eigenvalues generated asymptotically are considered to be valid
in all cases where normal modes exist.\par
Because this is a relaxation method reasonably good initial guesses
are required, both in the eigenfunction and the eigenvalue, in order
to accurately obtain an answer.  When the initial guesses were far
off from the actual solution the scheme admitted solutions belonging
to the continuous spectrum.  This occurs for all sets of boundary
condition but is the only solution possible for the case of fixed pressure conditions (see below).
Therefore we discuss the general features of the continuous spectrum in Section
\ref{fixed_pressure_conditions}.  Nonetheless, we find that it helps to avoid the continuous spectrum
if the initial eigenvalue guess is set so that ${\rm Im}(\omega) \neq 0$.
The possibility of
the solution jumping onto a random continuous mode solution is
satisfactorily bypassed in this way (also see below).
\par
In what follows we consider the discussion for each of the
three boundary conditions.  We finally note that the resulting
mathematical structure resulting from this asymptotic limit
is closely similar to the WKB analysis done in D05.  Whereas in
this study the small parameter is the horizontal wavenumber,
in D05 the
small parameter is the shear term $q \Omega_0 $ (denoted
by '$S$' in their equation 3).  In this sense, their asymptotic
form is valid for small values of the shear while ours is valid
for small values of $\alpha$.

\subsubsection{No normal-flow conditions}\label{section_wall_conditions}
The wall conditions, i.e.
that $  u_{_{\alpha\beta}} = 0$ at $x=0,1$,
becomes in terms of the pressure the requirement
\beq
\omega_e^2 u_1 = -i(\sigma_1\partial_x P_0 + 2\Omega_0\alpha_1 P_0) = 0,
\qquad {\rm at} \  x = 0,1,
\label{asymptotic_u1}
\eeq
at lowest order.
Implementing these conditions and a little algebra
gives a dispersion relation for $\omega_1$
\beqa
{\omega_1} &=& \alpha_1 q\Omega_0
\left(\frac{1}{2} \pm \frac{1}{2k_{_F}q}\Delta_{_F}^{1/2}\right),
\nonumber \\
\Delta_{_F} &=& 16 + k_{_F}^2 q^2 - \frac{8k_{_f}q}{ {\rm tanh}k_{_F}}
\label{wall_mode_dispersion}
\eeqa
As the dispersion relation clearly indicates, if $\Delta_{_F} < 0$ then
there appears a pair of complex modes, one which grows and one which decays.
When $\Delta_{_F} > 0$ there are two propagating modes oscillating with no overall
growth in amplitude.  The character of the stability is dictated only by the
two parameters $q$ and $k_{_F}$.
The limit where $N^2 \rightarrow 0$ (i.e. $k_{_F}\rightarrow \infty$)
reveals that
$
\omega_1 \rightarrow 0,q\alpha_1 \Omega_0.
$
%We recognize the latter of these solutions as the wall-mode discussed
%in the unstratified case. It
%is interesting that there is a neutral solution in this limit.
\par
\begin{figure}
\begin{center}
\leavevmode \epsfysize=6.85cm
\epsfbox{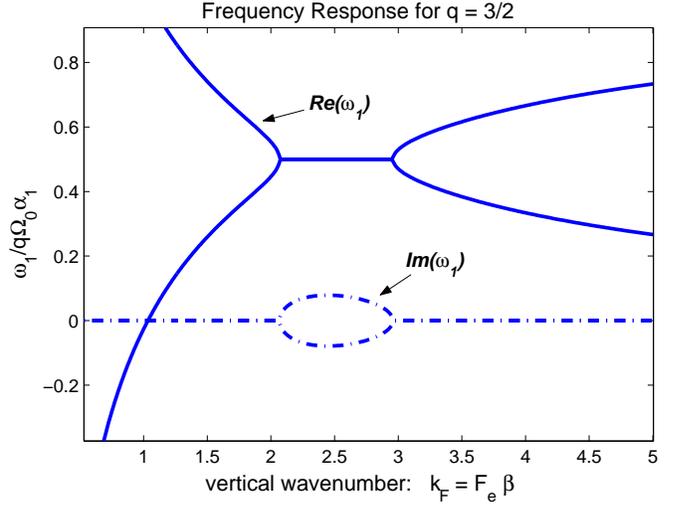}
\end{center}
\caption{{\small
Asymptotic
dispersion relation, $\omega_1$.  $q=3/2$
for channel flow disturbances.  The dashed curve is the growth/decay
while the solid line is the frequency.  The bifurcation into the
stable/unstable state begins and ends
near $k_F \sim 2.1, 3$ respectively.  In this range the frequency of
both modes are identical.
}}
\label{asymptotic_wallmode}
\end{figure}

The striking feature of this general
solution is that there exists a band of vertical wavenumbers for
which a stable/unstable solution exists.
In Figure \ref{asymptotic_wallmode} we plot this
dispersion curve for the case $q = 3/2$.  The plot shows a band in Froude-wavenumber
within which the stable/unstable pair
exists.  Recall that the Froude-wavenumber is really the vertical wavenumber
scaled by $F_e$.
The bifurcation into the stable/unstable pair occurs
when the frequencies of the two modes become the same.  In the
case depicted in the figure ($q=3/2$), when the
frequencies merge the instability emerges
near $k_F \sim 2.1$.  Until about $k_F \sim 3$, where
the instability vanishes, the frequencies remain the same.
\par
\begin{figure}
\begin{center}
\leavevmode \epsfysize=5.5cm
\epsfbox{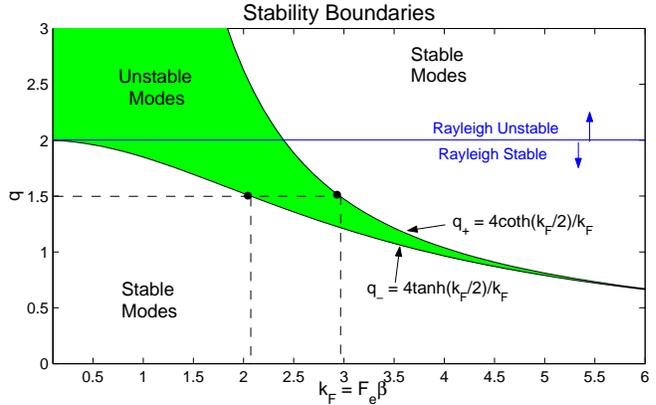}
\end{center}
\caption{{\small
Plotted is the stability boundary for channel flow disturbances
in the limit where $\alpha$ is small.  The shaded region
corresponds to the instability regime.  The bottom
portion of the instability region is bounded by the curve
$q_- = 4{\rm tanh}(F_e\beta/2)/F_e\beta$ while the upper
portion is bounded by
$q_+ = 4{\rm coth}(F_e\beta/2)/F_e\beta$.
The marginal stability
values for Keplerian flow, i.e. $q=3/2$, are labeled
by dashed lines.  The Rayleigh unstable line is
indicated, i.e. $q > 2$.
}}
\label{q_stability_boundary}
\end{figure}

The boundaries of the unstable band for general values of
$q$ may be inferred from the
expression for the growth rate: this means determining
the function $q_{\pm}(k_F)$ that satisfies
$\Delta_f(q,k_F) = 0$ as defined in (\ref{wall_mode_dispersion}).
The two functions are
$q_- = 4{\rm tanh(k_F/2)/k_F}$ and
$q_+ = 4{\rm coth(k_F/2)/k_F}$.  We see that the band structure
for the instability range persists until $q=2$, which happens to
also correspond to the Rayleigh instability line.
Beyond $q=2$ the instability range is bounded from below by zero
vertical wavenumber but it is still bounded from above by a finite
$\beta$.
 As the shear becomes weak,
the band of unstable modes gets correspondingly
thinner.  Note that this instability disappears
in the non-shearing limit, that is when $q\rightarrow 0$.
Figure \ref{q_stability_boundary} graphically summarizes these results.

\subsubsection{Fixed pressure conditions}\label{fixed_pressure_conditions}
The boundary conditions on the lowest order pressure conditions
becomes
\beq
P_{0}=0,
\eeq
at both boundaries $x=0,1$.  Given the form of the underlying equations
it turns out that there is no normal mode solution possible.
The numerical procedure admits solutions, however, these
are always modes {\em associated with the continuous spectrum} of
the linearized system.  They are not true normal modes in
the usual sense because they exhibit a discontinuity in some quantity:
here being in the horizontal velocity, $v$,
and manifesting explicitly as a step in the
quantity $\partial_x P$.  Discontinuities of this sort,
referred to sometimes as singular eigenfunctions, are typical
features of modes associated with a continuous spectrum
(Case, 1960, Balmforth \& Morrison, 1999).
In all cases, the location of the discontinuity is at some
value of $x=x_c$ which is the location of the {\em critical layer},
in other words, the place where
the quantity $\omega_1 - q\Omega_0 \alpha_1 x$ is zero.
This means (and this is verified numerically) for given values
of $\alpha$,$q$ and $\Omega_0$ there will be a continuum frequencies,
$\omega_c(\alpha,q,\Omega_0)$,
existing between $0$ and $q\Omega_0 \alpha$.
\par
Figure
\ref{continuous_spectrum_fig} displays examples of this continuum mode together
with
an analytic representation of the continuum mode (\ref{continuous_P_mode})
developed in Appendix \ref{continuous_mode}.

\subsubsection{Mixed conditions}\label{section_mixed_conditions}
\begin{figure}
\begin{center}
\leavevmode \epsfysize=10.0cm
\epsfbox{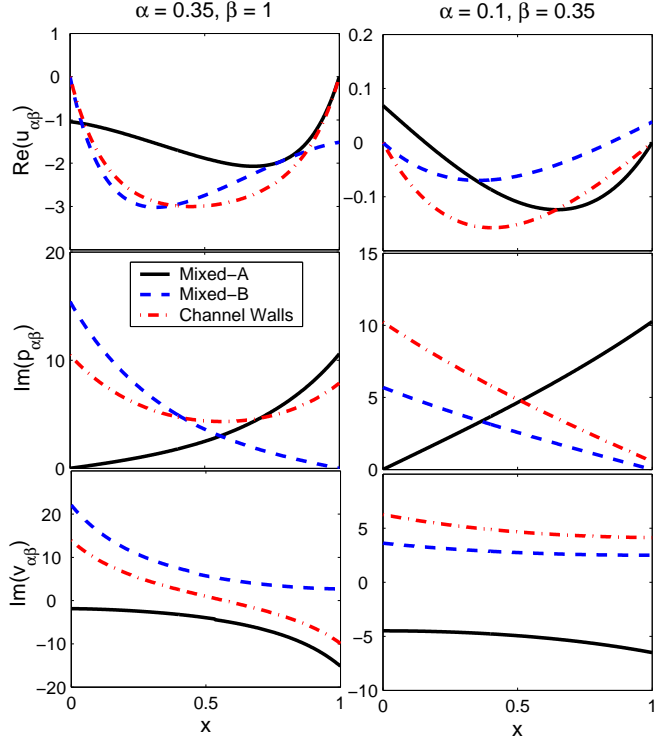}
\end{center}
\caption{{\small Assorted eigenfunctions for normal mode
solutions are plotted for $N = 0.4$.  In the first column
shows the solutions for the parameters $\alpha = 0.35, \beta = 1$
while the second column depicts solutions for
$\alpha = 0.1, \beta = 0.35$.  Each graph shows the behaviour
of the indicated quantity for the three sets of boundary
conditions:  channel walls (dashed-dot),
Mixed-A (dashed) and Mixed-B (solid).  Because for these
parameter values the resulting $\omega$ are real, the real parts
of the pressures and horizontal velocities are zero while the
imaginary parts of the radial velocities are zero.
}}
\label{eigenfunction_plots}
\end{figure}

We define terms by identifying {\em mixed-A} boundary conditions with
zero radial velocities at $x=0$ and zero pressure perturbation at $x=1$,
while
{\em mixed-B} boundary conditions indicate
zero pressure fluctuations at $x=0$ with zero radial
velocity perturbations at $x=1$.
Both boundary conditions yield single normal mode solutions
Consequently for mixed-A conditions the frequency
response is
\beq
\omega_1 = \Omega_0 q\alpha_1 - \frac{2\Omega_0 \alpha_1}{k_{_f}}
\tanh k_{_F},
\label{mixedA_omega}
\eeq
while for mixed-B conditions we have
\beq
\omega_1 = \frac{2\Omega_0 \alpha_1}{k_{_F}} \tanh k_{_F}.
\label{mixedB_omega}
\eeq
In Figure \ref{eigenfunction_plots}
we plot sample eigenfunctions for all boundary conditions we
considered in these sections, as well as comparisons between
the analytic and numerical solutions obtained.
\par
\begin{figure}
\begin{center}
\leavevmode \epsfysize=6.5cm
\epsfbox{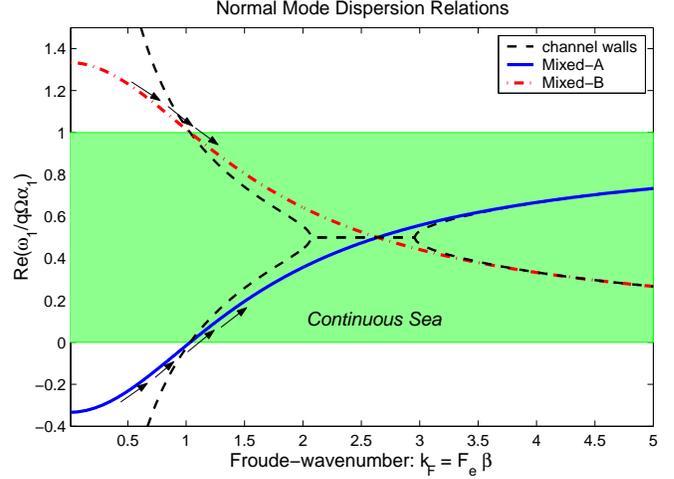}
\end{center}
\caption{{\small Normal mode dispersion plots
for the three sets of boundary conditions considered.
The frequency is shown scaled by $q\Omega_0 \alpha_1$.  Plotted
are the asymptotically determined dispersion relationships.  The shaded
region represents the continuous sea, i.e. modes associated with the
continuous spectrum.  Every point in the shaded region has a continuous spectrum
mode in it.  The strategy for following the discrete modes into the
continuous sea using the numerical method begins with locating a discrete
mode outside of the sea and then following the dispersion branch into the sea (as
indicated on the plot with arrows).
}}
\label{normalmode_dispersion_plots}
\end{figure}

It is important to mention that when discrete normal modes have frequencies which
sit in the continuous sea, i.e. when $0<\omega < q\Omega_0\alpha$ it becomes
challenging for the numerical method to not mistake it with a mode belonging to
the continuous spectra.  To circumvent this possible ambiguity (and
to properly numerically verify this limit) we
follow modes in $\beta$ starting initially with values of the discrete normal mode
frequency which is beyond the continuum  sea.  In this way,
discrete normal mode solutions are easily found
and, using these as a starting point, one
may  incrementally move into the realm where normal
modes exist
within the continuous sea.
This is depicted in Figure \ref{normalmode_dispersion_plots}.
In all boundary condition cases investigated, we successfully
trace the discrete mode spectrum.
\par

\par
\subsection{The QHSG approximation:
vertically varying $N^2$}
The asymptotic results of the last section hints toward
a tractable approach
in evaluating the generality of the SRI under a variety of
conditions.
The limit where the horizontal wavenumbers (i.e. $\alpha$)
are small suggests that there exists well-posed reduction of
the governing equations of motion (\ref{gssb_radial}-\ref{gssb_theta}).
With the azimuthal scales of disturbances scaling as $\order{1/\alpha}$
it followed that the temporal disturbances scale as $\order{\alpha}$ and, as
such, it implied that the radial and vertical velocities similarly
scale as $\order{\alpha}$ while the pressure, the temperature fluctuation
and the azimuthal velocities
scale as $\order{1}$.
We therefore propose that
when the horizontal scales are large compared to the corresponding
vertical and radial ones the
following scalings hold, in general, with respect to
quantities and operators of the system:
\beq
\partial_t, \partial_y, u, w \sim \order{\alpha}, \qquad
\partial_x, \partial_z, p, v, \theta \sim \order{1}.
\label{QHSG_scalings}
\eeq
This means that in the limit where the azimuthal scales are long,
we have the following effective reduction of the
nonlinear equations
of motion ({\ref{gssb_incompressible}-\ref{gssb_theta}),
\beqa
\partial_x u + \partial_y v + \partial_z w  &=& 0,
\label{QHSG_incompressible} \\
0 + \order{\alpha^2} &=& 2\Omega_0 v  -\frac{\partial_x p}{\bar\rho_b}, \label{QHSG_radial}\\
(\partial_t - q\Omega_0x \partial_y) v + {\bf u'}
\cdot\nabla v &=&
-(2-q)\Omega_0 u  -\frac{\partial_y p}{\bar\rho_b},
\label{QHSG_azimuthal}\\
0 + \order{\alpha^2} &=& -\frac{\partial_z p}{\bar\rho_b }
+ \frac{\theta g(z)\alpha_p}{\bar\rho_b} \label{QHSG_vertical}\\
(\partial_t - q\Omega_0x \partial_y)\theta' +
{\bf u'}\cdot \nabla\theta &=& - w \partial_z T_b.
\label{QHSG_theta}
\eeqa
The above set is similar to the quasi-geostrophic, quasi-hydrostatic
approximation used in the study of atmospheric flows
(e.g. Pedlosky, 1987, Salmon, 2002).  Whereas in the terrestrial
analog full quasi-geostrophy involves retaining the inertial terms
in the radial momentum equation, here they are absent (cf. \ref{QHSG_radial}),
and it is for
this reason we consider the above set of equations to be a sort of
{\em semi-geostrophic} limit.  The semi-geostrophic nature of this
set shares some similarities with the so-called
{\em elongated-vortices} equations derived
in Barranco et al. (2000).  The set presented
here differs from that work
in that the elongated-vortex equations do not make
the hydrostatic approximation as is a natural and necessary consequence here.
\par
The power in this reduced set of equations,
aside from exactly reproducing the asymptotic
limit explored in the previous section, is that it allows
one to investigate the effects that a position dependent
function of gravity and background state temperature gradient,
i.e. $g(z)$ and $ \partial_z T_b$, have on the SRI.  In this sense,
unlike the approximation utilized in Section \ref{BE_equations_constantN2},
we relax
the condition that $g$ and $\partial_z T_b$ are constants
and let them be general functions of $z$.  It therefore
means that the Brunt-Vaisaila frequency is now $z$-dependent.
\par
When specific
forms are considered here
we assume that these
quantities are simply proportional to $z$, that is to say,
\beq
g(z) = \Omega_0^2 z; \qquad \partial_z T_b = \tilde T_{zz} z.
\label{specific_g_and_Tb}
\eeq
The constant $\tilde T_{zz}$ sets the severity of the background
temperature gradient (see Section 1.1).
\par
We linearize the set
(\ref{QHSG_incompressible}-\ref{QHSG_theta}) about
the quiet state $u=v=w=\theta = 0$, $p =$ constant.
Disturbances are denoted with primes.
A little algebra
shows that these equations may be simplified into a single
one for the pressure perturbation:
\beq
(\partial_t - \Omega q x \partial_y)
\left[
\frac{\partial}
{\partial z} \frac{\omega_\epsilon^2}{N^2(z)}\frac{\partial p'}{\partial z}
+ \frac{\partial^2 p'}{\partial x^2}
\right] = 0,
\label{linear_potential_vorticity}
\eeq
where the $z$-dependent Brunt-Vaisaila frequency is given
as
\[
N^2(z) = \sfrac{1}{\bar \rho_b}
\alpha_p g \partial_z T_b =
\sfrac{1}{\bar \rho_b}
\Omega_0^2 \tilde T_{zz} \alpha_p z^2.
\]
This all also means that we can consider a $z$-dependent Froude number according to
\beq
F_{\epsilon}^2 = \frac{\omega_e^2}{N^2(z)} = \tilde F_{e}^2\frac{1}{z^2},
\eeq
because all variables and quantities have been non-dimensionalized,
the Froude-number scale $\tilde F_{e}$ should be considered in parallel
to the Froude-number $F_e$ treated in Section 2.2.\par
The expression inside the square brackets of
(\ref{linear_potential_vorticity}) looks analogous to
the {\em potential-vorticity} of atmospheric flows.
When $N^2(z)$ is a constant then the expression within the
square brackets exactly recovers the asymptotically valid
governing equation for the pressure perturbation
in (\ref{asymptotic_P0_eqn}).
\par
Separable solutions
are assumed of the form
\beq
p' = \Pi(z) P_0(x) e^{i\omega t + i\alpha y} + {\rm c.c.}
\label{separable_ansatz}
\eeq
and applied to the governing equation (\ref{linear_potential_vorticity}).
This results in two ODE's, one for
the vertical structure function and one for the radial structure function,
\beqa
\frac{\partial}
{\partial z} F^2_{\epsilon}\frac{\partial \Pi}{\partial z}
&=& - k_{_F}^2 \Pi, \label{Pi_eqn_QHSG}\\
\frac{\partial^2 P_0}{\partial x^2}  &=& k_{_F}^2 P_0
\label{P0_eqn_BE}
\eeqa
The separation constant for this procedure is $k_{_F}$.
We notice immediately that the equation for the radial structure function
is mathematically the equivalent to (\ref{asymptotic_P0_eqn}).
Since the boundary
conditions and associated relationships are identical in this
asymptotic limit, it follows
then the same stability properties that was determined
in Section \ref{BE_equations_constantN2}
carry over here to this particular example of a $z$-dependent
function of $N^2$ {\em if the allowed values of $k_{_F}$ are real}.
In remaining consistent with the terminology introduced in D05,  modes for which
$k_{_F}$ are real are referred hereafter as {\em exponential}, or as {\em e-modes},
since this describes the quality of the
radial structure function that results from solving (\ref{P0_eqn_BE}).
By contrast, modes in which the $k_{_F}$ are imaginary are referred to
as {\em oscillating}
or as {\em o-modes},
 (again see D05) and \emph{these in principle will have different stability
properties} than those determined for the e-modes.
\par
Thus the task that remains is to determine the allowed values of the
separation constant.  The analysis below indicates
that that the only types of disturbances
permitted for $N^2(z) \sim z^2$ are \emph{e-modes on a finite domain}.
\subsubsection{Exponential modes}
Though these may be artificial,
for the sake of simplicity and comparison we consider only the
vanishing of the normal velocities at the boundaries $z = \pm 1$
(e.g. Barranco \& Marcus, 2005).
When assuming the specific forms for $g$ and $\partial_z T_b$ as
in (\ref{specific_g_and_Tb}) we find two possible solution
forms to (\ref{Pi_eqn_QHSG})
\beq
\Pi(z) =
\left\{\begin{array}{c} \Pi_{(varicose)} \\
\Pi_{(sinuous)}
\end{array}
\right \}
=
\left\{\begin{array}{c}
%\Pi_{(varicose)}
z^{\frac{3}{2}}{\cal J}_{_{-\frac{3}{4}}}
\left(\frac{1}{2}\frac{k_{_F}}{\tilde F_e} z^2\right)
\\
%\Pi_{(sinuous)}
z^{\frac{3}{2}}{\cal J}_{_{\frac{3}{4}}}
\left(\frac{1}{2}\frac{k_{_F}}{\tilde F_e} z^2\right)
\end{array}
\right \}, \label{Pi_solutions} \eeq where the symbol ${\cal J}$
denotes the Bessel function of the first kind.
The expression $k_{_F}/\tilde F_e$ can be considered to be parallel to
the vertical wavenumber $\beta$ introduced and treated in Section 2.2.
A Taylor Series
expansion of these solutions near $z = 0$ verifies the even (odd)
symmetry of the varicose (sinuous) solutions. Given these inherent
symmetries we are left with the task of setting to zero $w'$ at
$z=1$ which, given the relationships between
(\ref{QHSG_vertical}-\ref{QHSG_theta}), is equivalent to setting
$\partial_z p' = 0$ there.  Given the general properties of Bessel
Functions (Abramowitz \& Stegun, 1972) the set of $k_{_F}$
values that satisfy the boundary conditions are always real. In
Figure \ref{vertical_eigenfunctions} we display the functions
$\Pi(z)$ for the first three values of the separation constant
$k_{_F}$.  It should be noted that the asymptotic expansion
of the solution forms presented
above are characterized by an amplitude function which
grows as $\sqrt z$ for $|z| \rightarrow \infty$.
%It is a genereric feature that the envelope of $\Pi(z)$
%increases with increasing $|z|$ because of its $\sqrt{|z|}$ dependence
%as $|z| \rightarrow \infty$.

\begin{figure}
\begin{center}
\leavevmode \epsfysize=7.25cm
\epsfbox{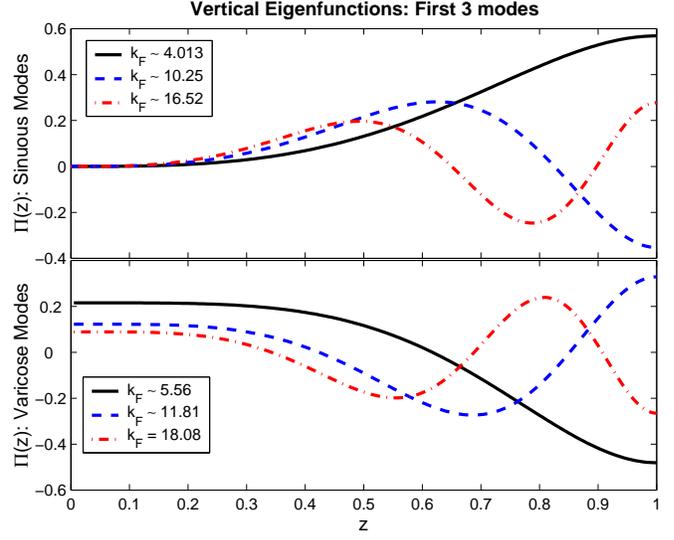}
\end{center}
\caption{{\small
The vertical eigenfunctions $\Pi(z)$ for e-modes are depicted for the first three
modes (varicose and sinuous) that satisfy the boundary condition
that the vertical velocity is zero at $z=\pm 1$.  Due to the inherent
symmetries in the modes, the behaviour of the function is depicted
for $z\ge 0$.
}}
\label{vertical_eigenfunctions}
\end{figure}

\subsubsection{Oscillating modes}\label{o-mode_analysis}
Consideration of o-modes starts by redefining the separation constant
$k_{_F}$ according to
\[
k_{_F} = i \kappa_{_F},
\]
where $\kappa_{_F}$ is real.  Therefore,
as in the previous section there are two independent solutions
to (\ref{Pi_eqn_QHSG}) with $k_{_F}$ so defined given by
\[ \sim \
z^{\frac{3}{2}} I_{_\frac{3}{4}}\left(\frac{1}{2}\frac{\kappa_{_F}}{\tilde F_e} z^2\right)
, \qquad
z^{\frac{3}{2}} I_{_{-\frac{3}{4}}}\left(\frac{1}{2}\frac{\kappa_{_F}}{\tilde F_e} z^2\right),
\]
respectively corresponding to sinuous and varicose disturbances.  However, Modified
Bessel Functions as these do not have zeros for real values of
the ratio $\kappa_{_F}/{\tilde F_e}$.
It means that it is not possible to satisfy
the same sort of boundary conditions considered in the previous section
(e.g. $w=0$ at $z = \pm 1$) which, in turn,
means that o-modes are not allowed
solutions on a finite vertical domain with no-vertical flow boundary conditions.\par
However, a consideration of the problem of (\ref{Pi_eqn_QHSG})
on an infinite domain (i.e. $z \rightarrow \pm\infty$)
where, instead, we require boundedness and asymptotic decay
of all quantities as $|z| \rightarrow \infty$,
suggests that the solution for $\Pi(z)$ could be the following,
\beq
\Pi(z)
=
z^{\frac{3}{2}}{\cal K}_{_{\frac{3}{4}}}
\left(\frac{1}{2}\frac{\kappa_{_F}}{\tilde F_e} z^2\right),
\label{Pi_solutions_oscillating}
 \eeq
 where
$ {\cal K}_{_\nu}
\left(x\right)$ is the Modified Bessel Function of the Second Kind
(Abramowitz \& Stegun, 1972).
 Indeed an asymptotic expansion of the leading order behavior
 of $ {\cal K}_{_{3/4}}
\left(z^2\right)$ in the large argument limit
 shows that it behaves like a Gaussian, i.e. like
 $\exp{-\sfrac{1}{2}\sfrac{\kappa_{_F}}{\tilde F_e} z^2}$, for large $z^2$.
 On its surface this behaviour seems to satisfy the requirements on the
 functions as $z \rightarrow \pm \infty$.  Although the above conclusion
 is correct for $z\rightarrow \infty$, to extend this conclusion
 as $z \rightarrow -\infty$ based on the above representation would be wrong.
 In fact, given the functional form (\ref{Pi_solutions_oscillating}), it becomes
 a matter of subtlety as to how one must cross the point $z=0$.  In Appendix \ref{KnuSubtlety}
it is shown that the behaviour of (\ref{Pi_solutions_oscillating})
\emph{for $z < 0$ is }
\beq
(-z)^{\frac{3}{2}}\left [ {\cal K}_{_{\frac{3}{4}}}
\left(\sfrac{1}{2}\sfrac{\kappa_{_F}}{\tilde F_e}(-z)^2\right)
- 2{\cal I}_{_{\frac{3}{4}}}
\left(\sfrac{1}{2}\sfrac{\kappa_{_F}}{\tilde F_e}(-z)^2\right)\right ].
\label{Pi_for_negative_z}
\eeq
Inspection reveals exponential divergence as $z \rightarrow -\infty$ since
${\cal I}_\nu(x)$ behaves exponentially as $x\rightarrow \infty$.
It appears that this analysis indicates that there are no bounded solutions possible
for $\Pi(z)$ and, consequently, it implies there are no o-modes permitted when
$N^2(z) \sim z^2$ in the context of the BE model.

\section{Large shearing box equations: the QHSG approximation and linearized
dynamics}\label{QHSG_LSB}
Given the clues revealed by using the QHSG for the BE, we consider the
same scalings expressed in
(\ref{QHSG_scalings}) and apply them to the full LSB equations (1-5).
The major departure here, of course, is that disturbances are now not incompressible.
One scaling relationship is to say that the density and pressure
variables are of comparable scale,
i.e. $\order{\rho} = \order{p} \sim 1$.
Therefore the QHSG reduction of the nonlinear LSB becomes
\beqa
& & (\partial_t - q\Omega_0x \partial_y)\rho + \nabla\cdot (\rho_b+\rho)\bf u'
= 0, \label{lsb_continuity_qhsg}\\
& &
%(\partial_t - q\Omega_0x \partial_y) u' + {\bf u'}
%\cdot\nabla u
0 = 2\Omega_0 v  -\frac{\partial_x p}{\rho_b + \rho}
+ \order{\alpha^2},
\label{lsb_radial}\\
%- \frac{\rho'g(z)}{\rho_b + \rho'}
& & (\partial_t - q\Omega_0x \partial_y) v' + {\bf u'}
\cdot\nabla u + (2-q)\Omega_0u = -\frac{\partial_y p}{\rho_b + \rho},
\label{lsb_azimuthal_qhsg}\\
& &
%(\partial_t - q\Omega_0x \partial_y) w' + {\bf u'}
%\cdot\nabla w
0 = -{\partial_z p}
- {\rho g(z)} + \order{\alpha^2},
\label{lsb_vertical_qhsg}\\
& & (\partial_t - q\Omega_0x \partial_y)\Sigma + {\bf u'}\cdot
\nabla(\Sigma_b + \Sigma) = 0,
\label{lsb_entropy_qhsg}
\eeqa
where we have introduced the basic state entropy $\Sigma_b$ and its
dynamically varying counterpart $\Sigma$ which are defined by
\beq
\Sigma_b = \ln{\frac{p_b}{\rho_b^\gamma}}, \qquad
\Sigma = \ln{\frac{1+\frac{p}{p_b}}{\left(1+\frac{\rho}{\rho_b}\right)^\gamma}},
\label{entropy_definition}
\eeq
where $\gamma$ is the the usual thermodynamic ratio of specific heats.
Linearizing
(\ref{lsb_continuity_qhsg}-\ref{lsb_entropy_qhsg}) and sorting
through the algebra (see Appendix \ref{lsb_qhsg})
leaves us with a single master equation for the pressure perturbation
\beq
(\partial_t - q\Omega_0 x\partial_y)\left[
\frac{\omega_\epsilon^2}{g}\partial_z p
+ \partial_z \frac{\omega_\epsilon^2}{N_{_\Sigma}^2}
\left(
\frac{g}{c^2}p + \partial_z p\right)
+\partial_x^2 p
\right] = 0.
\label{pv_lsb}
\eeq
The generalized Brunt-Vaisaila frequency is defined by
\beq
N^2_{_\Sigma} \equiv
\frac{g}{\gamma}\partial_z \ln \frac{p_b}{\rho_b^\gamma},
\label{N2_Sigma}
\eeq
while the (nondimensional) adiabatic soundspeed $c$ is defined by
\beq
c^2 \equiv \frac{\gamma p_b}{\rho_b}.
\eeq
(\ref{pv_lsb}) is the LSB equivalent, in this QHSG limit, of
a potential vorticity for a local section of a circumstellar disc.
Comparing this
equation for the potential-vorticity with
the analogous one for the BE in
(\ref{linear_potential_vorticity}) reveals some differences between them
being, namely,
\[
\frac{\omega_\epsilon^2}{g}\frac{\partial p}{\partial z} \ \ , \ \
\frac{\partial}{\partial z} \frac{\omega_\epsilon^2}{N_{_\Sigma}^2}
\frac{g}{c^2}p.
\]
The first of these is associated with the time rate of change of
the density fluctuation in the continuity
equation.  This is explicitly absent in the BE due to the
assumption of incompressibility.
The second of these is associated with the generalized entropy fluctuation
and is inversely proportional to the soundspeed.
This term is absent in the Boussinesq Equations because the assumption
of incompressibility is equivalent to the interpretation that {\em
the soundspeed is infinite}.
\par
Having
$N^2_{_\Sigma} < 0$ is equivalent to the Schwarzschild condition for
buoyant instability (Tassoul, 2000).
As before, we assume that the atmosphere is stable to buoyant oscillations
($N^2_{_\Sigma} > 0$).  However we must also say something about
the soundspeed $c$:  for the sake of this discussion we will assume
that it is a constant with respect to $z$, that is, we assume
the atmosphere is isothermal.
We proceed toward determining normal mode solutions of the expression
inside the square brackets of (\ref{pv_lsb}),
\beq
\frac{\omega_\epsilon^2}{g}\frac{\partial p}{\partial z}
+ \frac{\partial}{\partial z} \frac{\omega_\epsilon^2}{N_{_\Sigma}^2}
\left(
\frac{g}{c^2}p + \frac{\partial p}{\partial z}\right)
+\frac{\partial^2 p}{\partial x^2} = 0
\eeq
Assuming separable solutions of the form (\ref{separable_ansatz})
we find, once again, the following two problems to solve:
\beqa
\frac{\omega_\epsilon^2}{g}\frac{\partial \Pi}{\partial z}
+ \frac{\partial}{\partial z} \frac{\omega_\epsilon^2}{N_{_\Sigma}^2}
\left(
\frac{g}{c^2}\Pi + \frac{\partial \Pi}{\partial z}\right)
&=& -k_{_F}^2 \Pi,
\label{vertical_lsb_Pi_eqn} \\
\frac{\partial^2 P_0}{\partial x^2} &=& k_{_F}^2 P_0
\eeqa
The separation constant
$k_{_F}$ is the same as before.\par
Because the equation for $P_0$ is the same as in the BE model, cf. (\ref{P0_eqn_BE}),
it immediately follows that the same stability properties that was
determined for the BE apply here too if the set of
$k_{_F}$ values are all real (i.e. e-modes).
For this study we restrict
our attention to finite vertical domains (see below).
It means,
then, that the task that remains is to determine
the eigenvalues of $k_{_F}$ by seeking solutions of
(\ref{vertical_lsb_Pi_eqn}) subject to the boundary condition
that the vertical velocity vanishes at $z = \pm 1$.
In the
LSB, this condition amounts to setting
\beq
\frac{1}{N_{\Sigma}^2}\left(\frac{g}{c^2}\Pi + \frac{\partial \Pi}{\partial z}\right) = 0,
\eeq
at $z=\pm 1$. (see Appendix D).  Also, as before, we consider solutions to
$\Pi$ that are either sinuous or varicose.\par
We emphasize that we are restricting
our attention here to solutions on a finite $z$ domain.
This is because attention needed to treat such problems on an infinite domain
is challenging and it is, thus, outside the
scope of this current work (see Discussion).
\par
Aside from very special values
of the parameters, there
are no simple or analytically tractable
solutions to the ordinary differential equation posed
by (\ref{vertical_lsb_Pi_eqn}) \footnote{General solutions of this
equation are linear combinations of hypergeometric functions which
require numerical evaluation anyway.}.  Therefore we numerically
solve for this equation and $k_{_F}$ using a fourth-order variant of the NRK
scheme discussed in the previous section.  We use a grid of 300 points
which lets us determine solutions up to machine accuracy (i.e. an error
of less than $10^{-11})$.  The solutions were all normalized by setting $\Pi = 1$
at $z=1$.
We verify the robustness of the numerical
scheme by using it to solve the simpler equation (\ref{linear_potential_vorticity})
and comparing the numerically generated results against the exact solutions
(\ref{Pi_solutions}).
\par
There are two parameters that govern the solutions.  The first of these
is the scale measure, $\bar F_{_\Sigma}$, of the height-dependent Froude-number
\[
F_{_\Sigma}^2 = \frac{\omega_\epsilon^2}{N_{_\Sigma}^2} = \bar F_{_\Sigma}^2 z^2.
\]
Given that this atmosphere is
isothermal this Froude-number scale is measured by the parameter
\[
\bar F^2_{_\Sigma} = \frac{\omega_\epsilon^2 \gamma}{\Omega_0^2 (\gamma-1)}.
\]
For a medium dominated by molecular hydrogen $\gamma \approx 7/5$.  It means that
in a Keplerian flow  $\bar F^2_{_\Sigma} \approx 7/2$.
The second parameter is the relative measure of the vertical scale height of
the atmosphere defined by $H$ and given to be
\[
H^2 \equiv \frac{c^2}{\Omega_0^2} = \frac{\gamma {\cal R}_\mu \bar T}{\Omega_0^2},
\]
in which ${\cal R}_\mu$ is the non-dimensionalized gas constant for the given composition, $\bar T$
is the non-dimensionalized temperature of the atmosphere.  This quantity is essentially
the same as the classic $\epsilon$-parameter governing thin-disc theory
(Shakura and Sunyaev, 1973, Lynden-Bell and Pringle, 1974)
When $H$ is small, the atmosphere is very shallow and, consequently, cold.\par
The solutions that we scan all show that the $k_{_F}$ values are always real
indicating these are e-modes (cf. Section 2.3).  We were unable to find purely
imaginary solutions for $k_{_F}$ on this finite domain.
Thus it implies that the stability properties determined for e-modes
(i.e. Section 2) carry over to here too and that this system does not support o-modes
on this finite-domain.
%Lastly,
%colder atmospheres and larger Froude-numbers seem to imply larger wavenumbers.

\begin{figure}
\begin{center}
\leavevmode \epsfysize=10.2cm
\epsfbox{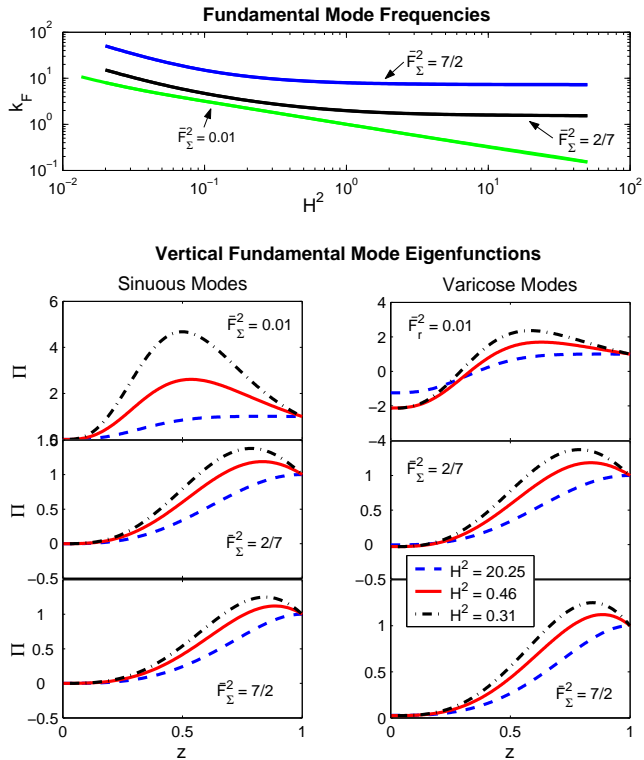}
\end{center}
\caption{{\small
Vertical wavenumbers and eigenfunctions for the fundamental mode
solutions of (\ref{vertical_lsb_Pi_eqn}) for selected values
of $H$ and $\bar F_r^2$.
}}
\label{lsb_plots}
\end{figure}

\section{Summary and Discussion}

\subsection{The QHSG and the persistence of the SRI with height dependence
in $N^2$}
We have achieved here an extension of the SRI to models which takes
into consideration the vertical structure of the physical environment.  One of the
departures taken here from previous work is to explicitly include
the effects of a vertically varying
Brunt-Vaisaila frequency.
The results of the previous sections shows
that the SRI, under channel-wall boundary conditions, persists unaltered
irrespective of the model equations considered (i.e. either the BE or the LSB)
or type of mode (i.e. e-mode or o-mode)
so long as one is in the inviscid-QHSG asymptotic limit.
This is not to say that if one relaxes the restriction of long horizontal
length scale disturbances (i.e. small $\alpha$) then this instability will not continue -
we merely mean to say that its existence under those conditions
remains open.  It does seem likely, however, given the pattern of the
presence of the SRI in D05, that it will do so also in the $\order{\alpha} \sim 1$
case too.\par
The advantage of the inviscid-QHSG approximation employed here is
that an analytical analysis of normal-modes is possible through
a separation of variables.  In general cases where both $g$ and the other
state variables like $\rho_b$ and $p_b$
are functions of the vertical coordinate $z$, the resulting linear equations
and mode structure
are generally non-separable.  This makes for assessing the normal-mode behaviour
of disturbances to be challenging (at best) although not impossible, as has
been shown here.  In this
sense it appears that this is a reasonable peek into situations where complicating
background structure in both the radial direction (here the shear) and vertical direction
(here gravity and other state variables) can be taken into account together.
\par
It is also worth noting
that in the inviscid QHSG limit of both the BE and LSB the radial
structure of the eigenmodes (be they either exponential or oscillatory) are unaffected by the vertical stratification: in
other words, the radial eigenfunctions are always the same.  As a result,
the stability criterion turns out to {\em be insensitive} to vertical stratification
in the state variables of the system.
\par
The QHSG limiting process is achieved by looking at disturbances with horizontal
length scales that are large compared to the other dimensions which, therefore,
implies that the associated time scale of disturbances
can scale with proportional smallness.  From the framework
of the the LSB equations, the implication is that one recovers the part of the dynamics
of inertio-gravity waves modified, to some extent, by the effect of weak compressibility
(dilatation) and a finite sound speed.  The inclusion of
these effects is not to say that some facet of acoustic disturbances are
recovered in this limit: this is precisely because the time scales associated with acoustics
are much shorter than the time scales explored here.  In this sense
this asymptotic limit
naturally filters out direct acoustic effects and preserves the dynamics of
disturbances that would usually
be associated with Rossby waves in geophysical flows (Pedlosky, 1987).
This is also not meant to imply that acoustic effects are unimportant (an issue that
is far from settled), it merely
means that this limit is effective at isolating the dynamics of these waves.
%This could mean that fully nonlinear investigations of the SRI
%with the assumption that $N^2$ is {\em constant} might yield results
%and implications that could prove to be very robust - they may be
%applicable for more realistic disc conditions in which $N^2$ does vary
%with height.
\par
\subsection{On the nature of e-modes and o-modes
with height dependent $N^2$}\label{exp_vs_osc_modes_disc}
D05 showed that there are two types of SRI modes which are
characterized by the quality of the mode's radial structure function: either
exponential or oscillatory.  Because both the vertical background temperature
gradient and component of gravity are constant D05 showed that
both e-modes and o-modes are supported for such a model atmosphere.  \par
For example, e-modes, which have vertical structure
functions which are sinusoidal with respect to the vertical coordinate,
are permitted in D05 even if
the atmosphere extends mathematically to infinity.  Because
$N^2$ is constant with respect to $z$,
the corresponding vertical structure functions have no envelope growth or
decay with respect to the vertical coordinate.  As such, such solutions
satisfy reasonable expectations of boundedness as the atmosphere extends indefinitely.\par
By contrast the analysis of
the QHSG limit of the BE equations, in which $N^2$ depends on $z$ quadratically,
shows that the resulting vertical
structure functions {\em for e-modes} to have envelope structures which grow ($\sim \sqrt {|z|}$).
This fact makes it impossible to
meaningfully impose boundary conditions or boundedness conditions
on solutions as $|z| \rightarrow \infty$.  The analysis performed here seems
to indicate that e-modes are restricted to BE systems involving finite
vertical domains with height dependent $N^2$.
\par
{\em O-modes} in model atmospheres with quadratic
dependence of $N^2$ with respect to $z$ are not allowed.
In atmospheres with constant $N^2$ o-modes are admitted
on account of the exponential decay of the vertical structure function.
For $N^2(z)\sim z^2$ (cf. \ref{o-mode_analysis})
it appears there is no way to construct bounded solutions
in the directions $z\rightarrow\pm\infty$ simultaneously.  It was
also demonstrated that o-modes are ruled out on finite domains.\par
The conclusions regarding the existence of o-modes is mainly based
on the analysis of the BE with $N^2(z)\sim z^2$ in the QHSG limit.
We showed also in Section 3 that similar conclusions seem to hold
for SRI e-modes and o-modes in the isothermal-QHSG limit of the LSB model set
when considered {\em only on
a finite domain}.  It is not entirely clear how the existence properties of SRI modes
are affected when: (i) the domain mathematically extends to infinity,
(ii) the soundspeed varies with height.  These questions are
reasonable points of departure for
further investigation.

\subsection{The absence of the SRI for non-reflecting boundaries}
The troubling aspect of this investigation is that when one considers
boundary conditions other than no-flow conditions on the radial
boundaries, the instability appears to vanish in the asymptotic limit
considered.  When the Lagrangian pressure is zero on both radial
boundaries the analysis predicts that there are no normal-modes with the time scales
assumed and, furthermore, there are only modes associated with the continuous
spectrum.
It is probably safe to conclude that for this type of boundary condition,
that there are no normal-modes whose frequencies have magnitudes on the
order of or greater than the small scaling parameter ($\alpha$, the horizontal
wavenumber) of the limit explored here.
%It is possible, of course, that there
%are normal modes with frequencies scales less than this but that has yet
%to be determined.
In situations where there is a mixture of no-flow conditions
on one boundary and no Lagrangian perturbations on the other, only
one normal-mode is admitted by the system which propagates
with no growth or decay.
\par
The circumstance encountered here shares a number of similarities with
the Eady problem of baroclinic instability
in geophysical shear flows (Eady 1949, Criminale \& Drazin, 1990).
Drazin and Reid (1994) show that the Eady problem is essentially equivalent to
the stability of inviscid plane Couette flow (pCf) subject to boundary conditions
where the pressure perturbations are fixed on the channel walls, instead
of the usual condition in which the normal velocities are set to zero.
As Case (1960) showed, inviscid pCf flows with no-normal flow boundary
conditions admit only
continuous spectrum modes and {\em no} discrete modes.  By contrast,
Criminale \& Drazin (1990) showed that the Eady problem
has,  in addition to the continuous spectrum,
a number of discrete modes present (which are possibly unstable under suitable
conditions of the disturbances) when disturbances in inviscid pCf flow
have fixed pressures at
the boundaries  (Criminale et al. 2003).
\par
For the SRI investigated here,
an entirely analogous situation occurs: the linear operator governing
the system
is mathematically equivalent to the one characterizing inviscid pCf, but,
the variables and stability characteristics are interchanged.
%As we noted earlier, in the limit explored here, the
%fundamental linear operator (\ref{asymptotic_P0_eqn},\ref{P0_eqn_BE}) is mathematically
%equivalent to the linear operator
%for two-dimensional plane Couette flow.
Whereas in inviscid pCf the operator operates on the radial velocity (wall-normal)
in the SRI case here it operates on the pressure perturbation.
Thus the inviscid pCf problem admits normal modes (no normal modes) for
fixed pressure (no normal flow) boundary conditions while
the SRI problem admits normal modes (no normal modes) for no normal flow
(fixed pressure) conditions.  Of course, both scenarios reveal the presence of
a continuous spectrum irrespective of the boundary conditions
employed.  It seems as though the inviscid pCf and SRI problems
have properties and stability characteristics that are interchanged.
\par
\subsection{Questions and a conjecture}
From a more physically motivated standpoint,
we have experimented with
these set of boundary conditions because they allow one
to exert some comparative control between conditions.  It is shown
in Appendix \ref{integral_section} that
disturbances
in the BE, subject to these
boundary conditions,
have a {\em total disturbance energy}, $E$, which evolves according to the exchange of
energy that takes place between the (Keplerian) shear and disturbance modes
via the Orr-Mechanism and measured by the
Reynolds Stress term, i.e. the RHS of (\ref{E_evolution_happybcs}).
Conditions other than these would cause there to
be some {\em net} work (positive or negative) to be performed on the layer
during the ensuing course
of the disturbances (Schmid \& Henningson, 2001).
\par
It is a puzzle, then, that in this QHSG limit there is an instability in the case
of no-normal flow conditions and none otherwise.
Although this is merely a conjecture, is it possible that the SRI
occurs because of the double reflecting boundary conditions?  The instability
shares many of the same properties of the acoustic instability uncovered by
Papaloizou \& Pringle (1984,1985), otherwise known as the PP instability (Li et al. 2000).
It is an instability of an acoustic
disturbance in a domain like this with
reflecting inner and outer walls in which
there exists a critical layer, sometimes referred
to as a {\em corotation radius} (Li et al., 2000).
The waves grow in a resonant fashion because the
reflecting walls, either one or both, allows for
repeated passages of the wave across the critical layer which
allows it to draw energy from the shear (Drury, 1985).
It  was found that the PP instability vanishes when the amplifying agent,
usually the second reflecting boundary,
is removed (Narayan, Goldreich \& Goodman, 1987).
\par
Like the PP instability, the SRI as determined in this
work are waves existing in a domain containing
a critical point along with reflecting boundaries.  When one of the boundaries
no longer reflects, there is no instability.
Although these are neither compressible modes nor two-dimensional
is it possible that the SRI arises in an analogous way due to the pathology
that afflicts the PP instability?  This is an open
question which should be clarified in future work.
\par
A clue towards this end might be found in the observation that there
exists a second energy integral, as developed in Appendix D,
involving a {\em total energy} expression $\cal F$ which says
something interesting.  The domain integral of the quantity $\cal F$ is conserved
under no-normal radial flow conditions whereas it is not for the others.
Perhaps the instability is related to this constraint placed on the
dynamics of the system?
\subsection{Flow two-dimensionalization and another conjecture}
We demonstrated in Section 2.1 that there exists a conserved quantity
($\Xi$) of
the general linearized
system of the BE
that is advected by the local basic shear.
This quantity, which looks like a potential vorticity, is conserved independent of
the QHSG asymptotic limit explored.  Its analog is implied to exist
in the for the LSB model equations as discussed by Tevzadze et al. (2003).
According to its definition,
(\ref{def_Xi}), $\Xi$ is composed of the vertical vorticity and a quantity representing
buoyancy motions driven by density fluctuations.  We also noted that in the
limit where the buoyancy oscillations become very strong the term associated
with it in $\Xi$ may become less important.  In this circumstance
it implies that the flow will take on a nearly two-dimensional character.
\par
In particular if the quantity $N^2$ becomes large then, according to
(\ref{Nsc_theta_lin})
 {\em one possible scaling between quantities in an initial value problem}
is to have $\Theta$ remain an order 1 quantity while the vertical velocity, $w$,
be $\order{N^{-2}}$.  If all other quantities remain correspondingly order 1,
that is to say if $u,v,P,\partial_x,\partial_y,\partial_z,\partial_t
\sim \order{1}$, then to lowest order it would imply that the disturbances are dynamically
in hydrostatic equilibrium and it would imply that the flow is nearly
two dimensional conserving its vertical component of vorticity (cf. \ref{conserved_Xi}-\ref{def_Xi}).
\par
Barranco and Marcus (2005) demonstrated,
in their shearing sheet simulations of a stratified fluid,
the appearance of coherent
vortical structures with vorticity vector pointing in the vertical direction.
When they manifested themselves, the anticyclonic vortices appear near the vertical boundaries
of the system, in other words, in that part of the atmosphere where the vertical
component of gravity is greatest in magnitude.  They also demonstrated the robustness
and persistence of these anticyclones by artificially removing the vortex structure
(after having developed) and replacing the flow field with noise.  They show
that the noisy spectrum quickly redeveloped into coherent anticyclone(s) much
as it is known to do so in two dimensional shear flows (e.g. Umurhan \& Regev, 2004).
This fact is consistent with the implications suggested by the advected conservation
of the linear quantity $\Xi$.  Of course, only a nonlinear reformulation and
reexamination of $\Xi$ can offer a more solid basis to any connection that
there may exist here.
\par
Is it possible that it is a generic feature
of stratified flow with a Couette shear profile and a vertically dependent
Brunt-Vaisaila frequency
(e.g. appropriate for a local representation of a circumstellar disk as here)
to behave two-dimensionally in substantial parts of the atmosphere significantly away
from the disk midplane, i.e. those regions
dominated by a large Brunt-Vaisaila frequency?

\section{Acknowledgements}
I thank the anonymous referee who made a critical suggestion
pertaining to the existence of oscillatory modes in this study.
I also thank Oded Regev for his suggestions and support
for this work.  I am also deeply
indebted to discussions with Professors Hugh Davies
and Eyal Heifetz
who helped me to cast
the asymptotic limits developed here
in terms of a quasi-geostrophic formalism.

\appendix

\section{An Integral Statement for the Boussinesq Equations}\label{integral_section}
It is instructive to consider
integrals of the system since they can help guide one
into deciding which boundary conditions to be used.  We begin
with by noticing that for the situations considered in here,
the functional forms relating $g$ and $\partial_z T_b$
are always constant multiplicative factors of each other (see
Sections 2.1 and 2.2).
Therefore we take the ratio of these two quantities to be {\em always}
a constant, that is,
\[
\partial_z T_b / g = {\rm constant},
\]
over the {\em full spatial} domain under consideration.
With this assumption in hand one may
(i) take the scalar product
of (\ref{gssb_radial}-\ref{gssb_vertical}) and $\rho_b {\bf u'}$ ,
(ii) multiply (\ref{gssb_theta}) by $ \theta g \alpha_p/\partial_z T_b$
and (iii) adding the results of (i) and (ii) together
and making use of the incompressibility condition
(\ref{gssb_incompressible})
to reveal
\beq
\left({\partial_t}-q\Omega_0 x \partial_y\right){\cal E} +
{\bf v}\cdot \nabla \left({\cal E} + p\right) = 0,
\label{E_eqn}
\eeq
where
\[  {\cal E}  \equiv
\frac{{\bar\rho_b\bf u'}^2}{2}
+ \frac{g\alpha_p}{\partial_z T_b}\frac{\theta^2}{2}.\]
Using
condition (\ref{gssb_incompressible}) once more,
we may integrate (\ref{E_eqn}) over the full spatial domain
to find,
\beq
\frac{d E}{dt} = -\int_{{\bf S}}{({\cal E}+p) {\bf u'}\cdot {\bf \hat n}} dS,
-\int_{{\bf V}}{qu'v' dV}
\label{E_eval}
\eeq
with
\[
E \equiv
\int_{{\bf V}}
{\left(\frac{\bar \rho_b{\bf u'}^2}{2}
+ \frac{g\alpha_p}{\partial_z T_b}\frac{\theta^2}{2}\right)
dV},
\]
in which
${\bf V}$ and ${\bf S}$ is the volume and surface-boundary of the domain
in which ${\bf \hat n}$ is the unit normal of the surface.
The above result is true in general for both linear
and nonlinear perturbations.
We interpret the quantities in ${\cal E}$ in the following way:
the term $\bar \rho_b{\bf u'}^2/{2}$
represents the kinetic energy in the disturbances
while the term
$g\alpha_p {\theta^2}/{2\partial_z T_b}$
represents the energy in thermal processes.
By definition ${\cal E}$ is zero in steady state, while
the steady pressure is constant, denoted by $\bar p$.\par
The global integral $E$, which we refer to as the
total energy in disturbances, can change due to
the influx of ${\cal E}$ across the boundaries,
through work done upon the system externally
as represented by
the boundary flux term $\int_{{\bf S}}p {\bf u'}\cdot {{\bf \hat n}} dS$,
and finally due to the
interaction with the background shear via the
Reynolds stress term
$-\int_{{\bf V}}{qu'v' dV}$
(for a discussion of this see Schmid \& Henningson, 2002).
\par
The general evaluation of
(\ref{E_eval}) may proceed once boundary conditions are specified.
As we have stated earlier, we will consider disturbances to be periodic
in the $y$ and $z$ directions (sinuous or varicose for the latter).
The radial boundary conditions and the motivation for
their choices
deserve some additional reflection.  We remind the reader that
one of the goals of this
study is to assess whether or not the SRI is an intrinsic
instability of the fluid and not some artifact
of boundary conditions.  One reasonable control
is to require that there is neither work done on the system
from outside nor there be any flux of energy across
the bounding walls.  This requirement requires that either
the normal velocities are zero on either of the two walls
or that (for linear disturbances only) the Lagrangian pressure perturbations are zero
at the two bounding surfaces.  A mixture of these
can also
affect the same outcome.
That is to say, for example, one could require that the normal
velocity at one bounding surface is zero while the Lagrangian
pressure perturbation is zero at the other surface.
Imposing these conditions therefore implies that
$E$ can change only due to the interactions of the
perturbations directly with the shear, in other words,
all such disturbances behave according to
\beq
\frac{dE}{dt} = -q\int_{{\bf V}} u' v' dV.
\label{E_evolution_happybcs}
\eeq
In this sense,
then, these solutions share some common property that
allows for some comparison.
\par

\section{Development of the continuous spectrum mode}\label{continuous_mode}
The discussion here largely follows the tack taken by Case (1960).
Beginning with (\ref{asymptotic_P0_eqn}) we can
say that
\beq
\left(\partial_x^2  - F_e^2\beta^2\right) P_0
= 0,
\label{simple_operator}
\eeq
is true for $x \neq x_c$ where
\[
x_c = \frac{\omega_1}{q\Omega_0 \alpha_1}.
\]
Taking $P_0^{\mp}$ to denote the solution of
(\ref{simple_operator}) to the left and right (respectively)
of $x_c$, and, assuming zero pressure conditions at $x=0,1$
we have that
\beq
P_0^{-} = A^{-}\sinh\left[k_{_F} x\right], \qquad
P_0^{+} = A^{+}\sinh\left[k_{_F} (1-x)\right]
\eeq
Then, to enforce continuity of the pressure at $x=x_c$
we see that
\[
A^{-} =  A^{+}
\frac{\sinh\left[k_{_F} (1-x_c)\right]}
{\sinh\left[k_{_F} x_c\right]}.
\]
Once some normalization is specified, that is, a value of
$A^{+}$ is assumed, the solution is complete.
For numerically generated solutions, we set $P(x=0.99) = 0.01$.
In summary,
then, we have the continuous mode pressure eigenfunction,
$P^{(c)}_0$, is
\beq
P_0^{(c)} =
A^{+}
\left\{
\begin{array}{lr}
\frac{\sinh\left[k_{_F} (1-x_c)\right]}
{\sinh\left[k_{_F} x_c\right]}
\sinh\left[k_{_F} x\right] & 0<x<x_c,  \\
\sinh\left[k_{_F} (1-x)\right] & x_c<x<1.
\end{array}
\right. ,
\label{continuous_P_mode}
\eeq
in which $A^{+} = 0.01/\sinh{[0.99\times k_{_F}]}$.
Note that it means that any value of $\omega_1$ which satisfies
the requirement $ 0 < x_c < 1$ is an allowed solution.  This is
the nature of the continuous spectrum (Schmid \& Henningson, 2001).
Also, the mode associated with the
continuous spectrum is not technically defined at $x=x_c$.
\begin{figure}
\begin{center}
\leavevmode \epsfysize=10.0cm
\epsfbox{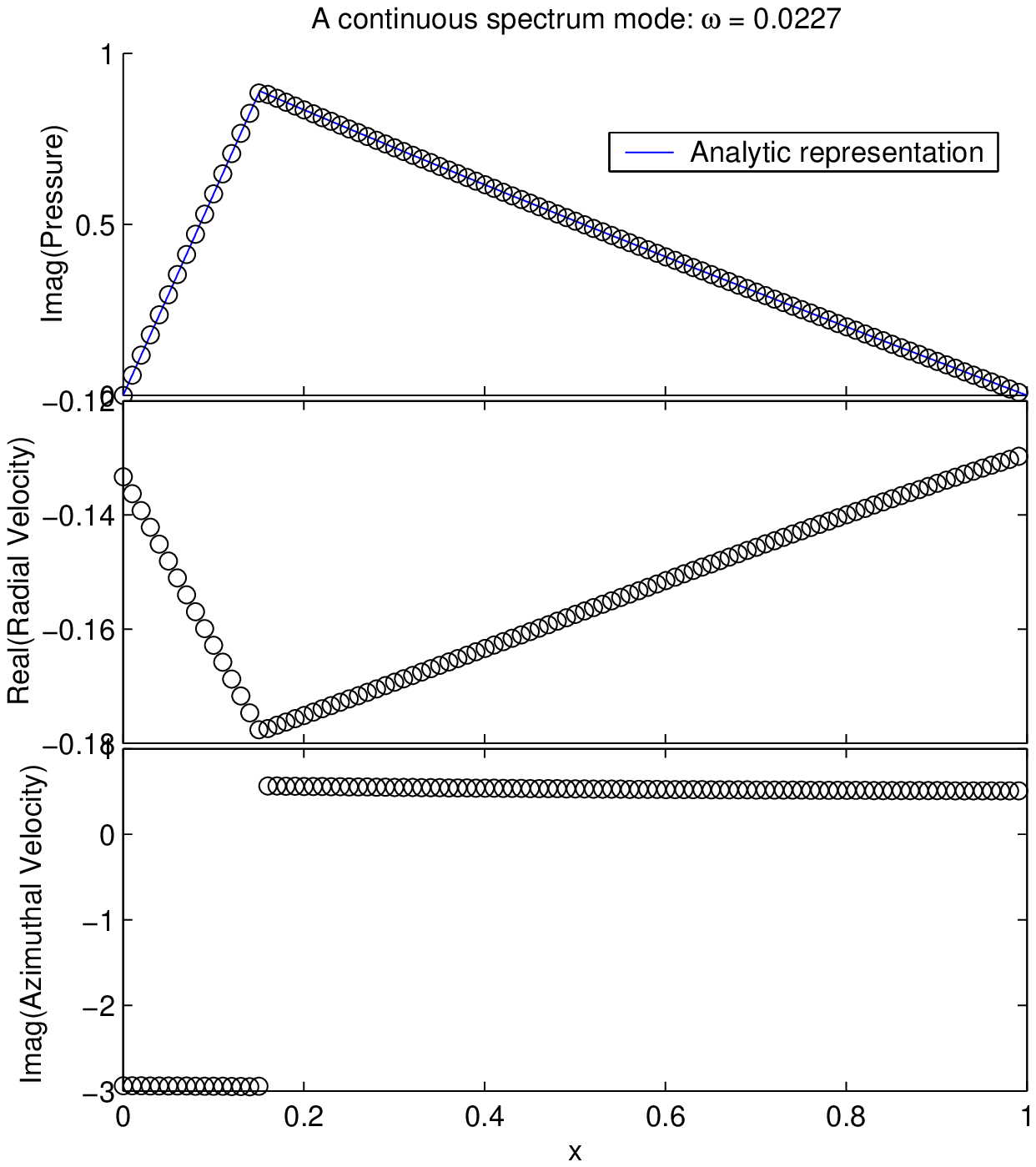}
\end{center}
\caption{{\small
An example of the numerically generated continuous spectrum mode for
fixed pressure conditions with $\alpha = 0.1, q= 3/2, \Omega_0 = 1$.
This particular mode has a frequency $\omega \approx 0.023$.  Every
seventh data point is plotted for the sake of visual clarity.
The analytic representation of the eigenmode, i.e (\ref{continuous_P_mode}),
is shown in the top graph for pressure
only.  In all cases the pressure eigenfunction
is normalized to be $0.01$ at $x = 0.99$.  The jump in the horizontal
(azimuthal) velocity is plainly evident.
}}
\label{continuous_spectrum_fig}
\end{figure}

\section{Subtlety in the $K_\nu$ oscillating mode solution}\label{KnuSubtlety}
 A closer analysis of
 (\ref{Pi_solutions_oscillating}) reveals that it
 is everywhere analytic and, therefore, an entire function along the real $z$
 axis, despite the presence of a branch point at $z=0$ for the $K_{\nu}$
 Bessel function.  It means, therefore, that in order to express the
 nature of this function as you one passes through $z=0$ one must be very careful
 about the relative phases that are incurred by crossing the $z=0$ point.
 To be more specific, let us define
 \[
  \zeta = \left | \frac{\kappa_{_F}}{\tilde F_{e}} \right |^{1/2} z
  \]
  and rewrite the solution
  (\ref{Pi_solutions_oscillating}) according to its composition of Modified
  Bessel Functions of the first kind,
\[
\Pi(z) \sim
\zeta^{\frac{3}{2}}\left [ {\cal I}_{_{\frac{3}{4}}}
\left(\sfrac{1}{2}\zeta^2\right)
-
{\cal I}_{_{-\frac{3}{4}}}
\left(\sfrac{1}{2}\zeta^2\right),
\right]
\]
which, according to the series representation of $I_{_\nu}$
would be
\[
= \zeta^{\frac{3}{2}}\left [\zeta^{\frac{3}{2}}\Upsilon(3/4,\zeta)
- \zeta^{-\frac{3}{2}}\Upsilon(-3/4,\zeta)\right ]
\]
where
\[
\Upsilon(\nu,\zeta) = \left(\sfrac{1}{4}\right)^{\nu}\sum_{k=0}^{\infty}
\frac{\left(\sfrac{1}{16}\zeta^4\right)^k}{k!\Gamma(\nu+k+1)},
\]
(Abramowitz \& Stegun, 1972).
The function $\Upsilon$ is symmetric with respect to the reflections
$\zeta \rightarrow -\zeta$.  However, the pre-factors appearing in
the above expressions imply that one must be very careful in interpreting
the function as $\zeta$ crosses over zero.  This is best illustrated by
considering the behaviour
of (\ref{Pi_solutions_oscillating}) for $\zeta < 0$.  Expressed
initially without restriction of $\zeta$
(\ref{Pi_solutions_oscillating}) is,
\beqa & = &
\zeta^{3}\Upsilon(3/4,\zeta)
- \Upsilon(-3/4,\zeta) \nonumber
\eeqa
Then if we restrict attention to $\zeta < 0$
by defining the variable $s = -\zeta$ and restricting attention
to $s > 0$ we see that the above becomes
\beqa& = &
-s^{3}\Upsilon(3/4,s)
- \Upsilon(-3/4,s) \nonumber \\
& = &
s^{\frac{3}{2}}\left [-s^{\frac{3}{2}}\Upsilon(3/4,s)
- s^{-\frac{3}{2}}\Upsilon(-3/4,s)\right ] \nonumber \\
& = &
s^{\frac{3}{2}}\left [-{\cal I}_{_{\frac{3}{4}}}
\left(\sfrac{1}{2}s^2\right)
- {\cal I}_{_{-\frac{3}{4}}}
\left(\sfrac{1}{2}s^2\right)\right ] \nonumber \\
& = &
s^{\frac{3}{2}}\left [-2 {\cal I}_{_{\frac{3}{4}}}
\left(\sfrac{1}{2}s^2\right)
+ {\cal K}_{_{\frac{3}{4}}}
\left(\sfrac{1}{2}s^2\right)\right ] \nonumber
\eeqa
rewriting the above we see that the solution $\Pi(z)$ for
$z< 0$ becomes as it is expressed in the text (\ref{Pi_for_negative_z}).

\par
\section{QHSG Linearization of the LSB}\label{lsb_qhsg}
We linearize (\ref{lsb_continuity_qhsg}-\ref{lsb_entropy_qhsg}).
It now means that $\rho'$ and $p'$ are the linearized density and
pressure fluctuations.
The resulting equations become
\beqa
& & (\partial_t - q\Omega_0x \partial_y)\rho' +
\partial_x m_u + \partial_y m_v + \partial_z m_w
= 0, \label{linear_lsb_continuity_qhsg}\\
& &
%(\partial_t - q\Omega_0x \partial_y) m_u + {\bf u'}
%\cdot\nabla u
0 = 2\Omega_0 m_v  -\partial_x p',
\label{linear_lsb_radial_qhsg}\\
%- \frac{\rho'g(z)}{\rho_b + \rho'}
& & (\partial_t - q\Omega_0x \partial_y) m_v
+ (2-q)\Omega_0 m_u = -\partial_y p',
\label{linear_lsb_azimuthal_qhsg}\\
& &
%(\partial_t - q\Omega_0x \partial_y) w' + {\bf u'}
%\cdot\nabla w
0 = -{\partial_z p'}
- {\rho' g(z)},
\label{linear_lsb_vertical_qhsg}\\
& & (\partial_t - q\Omega_0x \partial_y)\rho_b \Sigma' + m_w \partial_z S_b = 0,
\label{linear_lsb_entropy_qhsg}
\eeqa
where for the sake of compact notation we have introduced the perturbed
momentum fluxes
\[
m_u \equiv \rho_b u', \qquad
m_v \equiv \rho_b v', \qquad
m_w \equiv \rho_b w',
\]
and the perturbed entropy
\[
\Sigma' = \frac{\gamma}{\rho_b} \left(\frac{p'}{c^2} - \rho'\right).
\]
Operating on (\ref{linear_lsb_azimuthal_qhsg}) with $\partial_x$ and then
making use of (\ref{linear_lsb_continuity_qhsg}) reveals
\[
(\partial_t - q\Omega_0x \partial_y) \left[\partial_x m_v  - (2-q)\Omega_0 \rho' \right]
=
(2-q)\Omega_0 \partial_z m_w.
\]
This is followed by multiplying (\ref{linear_lsb_entropy_qhsg}) by
$\Omega_0(2-q)/\partial_z S_b$ and then operating on the result with
$\partial_z$.  This gives
\[
(\partial_t - q\Omega_0x \partial_y)
\frac{\partial}{\partial z} \Omega_0 (2-q)\frac{\rho_b \Sigma'}{\partial_z S_b}
= -(2-q)\Omega_0 \partial_z m_w.
\]
Adding these two equations gives
\[
(\partial_t - q\Omega_0x \partial_y)\left[
\partial_x m_v  - (2-q)\Omega_0 \rho' +
\frac{\partial}{\partial z} \Omega_0 (2-q)\frac{\rho_b \Sigma'}{\partial_z S_b}
\right]
=0.
\]
Given the definition of $\Sigma'$, the hydrostatic relationship (\ref{linear_lsb_vertical_qhsg}),
and the radial geostrophic balance (\ref{linear_lsb_radial_qhsg}) recovers
(\ref{pv_lsb}).

\section{A Second Integral Statement of the Boussinesq Equations}\label{full_energy_integral}
Following the steps in Section
\ref{integral_section} one may generate a second energy integral.
The dynamical equations (\ref{gssb_incompressible}-\ref{gssb_theta})
describe the evolution
of the both the disturbance velocities, i.e.
the
velocity fluctuations over and above the steady state Keplerian velocity,
and the temperature fluctuations.
We denote
the {\em total} velocity of disturbances \textit{in the frame of the shearing box} as
$\bf U$ and given to be
\beq
{\bf U} \equiv -q\Omega_0 x {\bf{ \hat y}} + {\bf u'} = \{u', v'-q\Omega_0 x, w' \}
\eeq
As such the governing equations of motion (\ref{gssb_radial}-\ref{gssb_theta})
are more concisely written in vector form as
\beqa
\partial_t{\bf U} + {\bf U}\cdot \nabla {\bf U} &=& -
\sfrac{1}{\bar \rho_b}
\nabla p  \nonumber  \\
&+& 2\Omega_0 {\bf{\hat z}}\times({\bf U} + q\Omega_0 x
{\bf{\hat y}}) + \sfrac{1}{\bar \rho_b}g\alpha_p\theta
{\bf{\hat z}} \label{total_velocity_eqn} \\
\partial_t{\theta} + {\bf U}\cdot \nabla {\bf \theta} &=& -w\partial_z T_b
\label{total_theta_eqn}
\\
\nabla\cdot{\bf U} &=& 0 \label{total_incompressibility}.
\eeqa
As we have posited $\partial_z T_b$ and $g$
multiplicative factors of each other
over the {\em full spatial} domain under consideration.
With this assumption in hand one may
(i) multiply (\ref{total_velocity_eqn}) by $\rho_b {\bf v}$ ,
(ii) multiply (\ref{total_theta_eqn}) by $ \theta g \alpha_p/\partial_z T_b$
and (iii) adding the results of (i) and (ii) together to yield
\beq
{\partial_t}{\cal F} +
{\bf U}\cdot \nabla \left({\cal F} + p\right) = 0,
\label{F_eqn}
\eeq
where
\[  {\cal F}  \equiv
\frac{{\bar\rho_b\bf U}^2}{2}
+ \frac{g\alpha_p}{\partial_z T_b}\frac{\theta^2}{2}
- q\Omega_0^2 x^2. \]
With use of the incompressibility
condition (\ref{total_incompressibility})
we may integrate (\ref{F_eqn}) over the full spatial domain
to find,
\beq
\frac{d\Phi}{dt} = -\int_{{\bf S}}{({\cal F}+p) {\bf U}\cdot {\bf \hat n}} dS,
\label{F_eval}
\eeq
with
\[
\Phi \equiv \int_{{\bf V}}{\cal F}dV
=
\int_{{\bf V}}
{\left(\frac{\bar \rho_b{\bf U}^2}{2}
+ \frac{g\alpha_p}{\partial_z T_b}\frac{\theta^2}{2}
- q\Omega_0^2 x^2\right)dV},
\]
in which
${\bf V}$ and ${\bf S}$ are as they were defined before.
We interpret the quantities in ${\cal F}$ in the following way:
the term $\bar \rho_b{\bf U}^2/{2}$
represents the kinetic energy,
the term $-q\Omega_0^2 x^2$ is like a potential energy
and
$g\alpha_p {\theta^2}/{2\partial_z T_b}$
represents the energy in thermal processes.
The global integral $\Phi$ can change due to
the influx of ${\cal F}$ across the dynamically undulating boundaries as
well as through the work done upon the system externally
as represented by
the boundary flux term $\int_{{\bf S}}p {\bf U}\cdot {{\bf \hat n}} dS$.
\par
The point of this exercise is to note that only for no-normal flow boundary
conditions does $\Phi$ remain fixed for disturbances.  The other conditions,
like fixing the Lagrangian pressure, can cause $\Phi$ to vary over the course
of its evolution.  This is because although $\bar p$ may be constant
in steady state, the total quantity ${\cal F}$ is not constant in steady state
where a simple inspection of its definition clearly reveals.   By fixing only
the Lagrangian pressure fluctuation, the otherwise moving boundary
can allow ${\cal F}$ to seep in and out of the domain.
\par
Perhaps, then, the reason for the existence of the instability
under no-normal flow boundary conditions arises because of this preserved
property of the disturbances.  The reflection property of the boundaries perhaps
traps energy in a way that causes growth to be encouraged.  In this case, the
energy of the disturbances must come from the energy contained in the background
shear state and because there is an overall trapping of the energy, a runaway
extraction processes takes place - ironically, leaving the total energy
budget , $\Phi$, fixed over the course of the evolution.

\end{document}